\documentclass[aps,prd,reprint,superscriptaddress,nofootinbib,longbibliography,floatfix]{revtex4-2}
\usepackage[T1]{fontenc}
\usepackage{lmodern}
\usepackage{amsmath,amssymb,amsfonts,bm}
\usepackage{graphicx}
\usepackage{xcolor}
\usepackage{hyperref}
\hypersetup{colorlinks=true,linkcolor=blue,citecolor=blue,urlcolor=blue}
\allowdisplaybreaks

\newcommand{\Mpl}{M_{\rm Pl}}
\newcommand{\dd}{\mathrm{d}}
\newcommand{\e}{\mathrm{e}}
\newcommand{\ii}{\mathrm{i}}
\newcommand{\cH}{\mathcal{H}}
\newcommand{\Pg}{\mathcal{P}}

\newcommand{\AL}{A_L}
\newcommand{\FD}{\mathcal{F}_{\Delta N}}
\newcommand{\AD}{\mathcal{A}_{\Delta N}}

\newcommand{\CD}{\mathcal{C}_{\Delta N}}
\newcommand{\RD}{\mathcal{R}_{\Delta N}}

\newcommand{\rDM}{r_A}

\begin{document}

\title{Finite-duration non-slow-roll effects in inflationary production of ultralight vector dark matter}

\author{Imtiaz Khan}
\email{ikhanphys1993@gmail.com}
\affiliation{Department of Physics, Zhejiang Normal University, Jinhua, Zhejiang 321004, China}
\affiliation{Research Center of Astrophysics and Cosmology, Khazar University, Baku, AZ1096, 41 Mehseti Street, Azerbaijan}

\author{Niamat Ullah}
\email{Niamat.ullah@buitms.edu.pk}
\affiliation{Department of Physics, Balochistan University of Information Technology, Engineering and Management Sciences (BUITEMS), Quetta, Balochistan, Pakistan.}

\author{Chengxun Yuan}
   \email{yuancx@hit.edu.cn}
\affiliation{School of Physics, Harbin Institute of Technology, Harbin 150001, People’s Republic of China}

\author{G. Mustafa}
\email{gmustafa3828@gmail.com}
\affiliation{Department of Physics, Zhejiang Normal University, Jinhua, Zhejiang 321004, China}

\author{Farruh~Atamurotov}
\email{atamurotov@yahoo.com}
\affiliation{Kimyo International University in Tashkent, Shota Rustaveli str. 156, Tashkent 100121, Uzbekistan}
\affiliation{National University of Uzbekistan, Tashkent 100174, Uzbekistan}

 \author{Ahmadjon~Abdujabbarov}
	\email{ahmadjon@astrin.uz}
\affiliation{School of Physics, Harbin Institute of Technology, Harbin 150001, People’s Republic of China}
\affiliation{University of Tashkent for Applied Sciences, Str. Gavhar 1, Tashkent 100149, Uzbekistan}

\begin{abstract}

Inflationary production of ultralight vector dark matter can be amplified by a brief non-slow-roll phase, yet instantaneous mass transitions generate ultraviolet ringing and leave finite-time predictions unresolved. We replace the sharp transition with a smooth profile and derive the momentum-response kernel of the longitudinal mode. Finite duration preserves the infrared enhancement, suppresses modes that resolve the transition, and introduces a profile-dependent correction to the relic abundance. After fixing the dark matter abundance, the induced gravitational-wave signal retains an independent transverse-traceless source dependence, so abundance rescaling alone leaves residual profile information. Numerical mode evolution and radiation-era tensor integration support the finite-width treatment. Transition duration therefore links microscopic inflationary dynamics to ultralight vector abundance and its low-frequency gravitational-wave signal.
\end{abstract}
\maketitle

\section{Introduction}

Dark matter has firm astrophysical and cosmological support, while its particle identity remains unresolved \cite{BertoneHooperSilk2005,Feng2010}. Inflationary particle production offers a direct origin mechanism: accelerated expansion can populate hidden-sector fields even for extremely weak couplings to Standard Model particles \cite{Parker1969,Ford1987}. The resulting abundance then probes inflationary dynamics and quantum fields in curved spacetime.

For scalar dark matter candidates, inflationary production is commonly associated with mechanisms such as vacuum misalignment, spectator field fluctuations, and isocurvature constraints arising from cosmic microwave background observations \cite{Marsh2016,Hui2017,Hui2021,Ijaz:2023cvc,Ijaz:2024zma,GarciaPierreVerner2023}. Massive vector fields have a different inflationary degree of freedom structure. In the light mass regime the transverse polarizations remain approximately conformal, while a gauge invariant mass makes the longitudinal polarization dynamical. The origin of that mass therefore determines which fields participate in the production problem. An Abelian Stueckelberg realization preserves the gauge symmetry and contains the three massive vector polarizations \cite{RueggRuizAltaba2004}. A Higgs realization also contains a radial scalar and its evolution can modify the produced vector spectrum and abundance when that scalar remains dynamical \cite{SatoTakahashiYamada2022,RediTesi2022}. Specifying the mass generation regime is consequently part of the inflationary calculation itself, especially when the mass history is resolved over a finite interval. The vector spectrum in the Stueckelberg regime acquires a blue infrared behavior and develops a characteristic peak near the scale that later becomes nonrelativistic during radiation domination. This structure suppresses large scale isocurvature perturbations and provides a viable mechanism for generating cosmologically relevant dark matter abundances \cite{GrahamMardonRajendran2016}. Broader phenomenological aspects of hidden photon and dark vector scenarios, including laboratory searches and astrophysical constraints, have been extensively studied in Refs.~\cite{Arias2012,CaputoMillar2021}.

The theoretical importance of inflationary vector dark matter production has motivated significant developments in recent years. Alternative production channels include purely gravitational generation, dark-photon amplification during inflation, charged-inflaton and symmetry-breaking realizations, as well as coherent oscillation and time-dependent vector mass scenarios \cite{EmaNakayamaTang2019,AhmedGrzadkowskiSocha2020,KolbLong2021,NakaiNambaWang2020,SalehianGorjiFirouzjahiMukohyama2021,FirouzjahiGorjiMukohyamaSalehian2021,Nakayama2019,Franciolini:2026cps,Nakayama2020Constraint,KitajimaNakayama2023,KanetaLeeLeeYi2023}. Additional extensions involving nonminimal curvature couplings further modify the effective mass evolution during inflation and substantially enlarge the allowed parameter space for viable dark matter production \cite{OzsoyTasinato2024,CapanelliJenksKolbMcDonough2024}. More generally, recent studies of controlled inflationary valleys, dilaton-flattened axion sectors, and heavy-field induced background deformations illustrate how microscopic inflationary dynamics can generate nontrivial time-dependent effective masses through transient departures from conventional slow-roll evolution \cite{Pirzada:2026jml,Pirzada:2026sle}. Complementary gradient-expansion approaches have further improved the connection between inflationary mode evolution and the final dark matter abundance \cite{LysenkoSobolVilchinskii2026}.

An additional phenomenological consequence of inflationary vector production is the generation of secondary gravitational waves. The enhanced longitudinal spectrum acts as a source for tensor perturbations at second order during radiation domination, allowing gravitational wave observations to probe vector dark sectors even in the absence of direct couplings to visible matter \cite{MarriottBestPelosoTasinato2025}. Moreover, the phenomenology of ultralight vector dark matter differs substantially from scalar fuzzy dark matter. Distinctive signatures may arise through anisotropic stress, polarization-dependent effects, coherent metric oscillations, modifications of Boltzmann evolution, and pulsar timing observables \cite{ChaseLeizerovichNacirLandau2024,ChaseCLASS2025,DrorWei2025,NomuraOmiyaTanaka2025}. These developments make inflationary vector dark matter a particularly interesting framework for connecting early-Universe physics with future observational probes.

The ultralight regime exposes a specific dynamical question. In the original slow roll mechanism the relic abundance is strongly correlated with the inflationary Hubble scale and the late vector mass \cite{GrahamMardonRajendran2016}. Rapid mass excursion calculations show that a short departure from slow roll can enhance the longitudinal mode and move the viable abundance branch toward much lighter vectors \cite{LaRosaTasinato2025}. That result is obtained in the sharp transition limit, where the event is characterized by its net logarithmic mass impulse. Higgs realizations address the gauge invariant origin of a vector mass and show that an active radial scalar can change the vector production history \cite{SatoTakahashiYamada2022,RediTesi2022}. These results leave a specific intermediate problem for the three polarization vector system. A physical mass excursion generated by an inflationary background lasts a finite interval, so the integrated impulse determines only the zero frequency response. Modes whose oscillation time is comparable to the duration accumulate phase while the mass evolves and resolve the temporal profile and its differentiability. The relic abundance weights one filtered vector spectrum while the induced tensor source convolves two filtered spectra. Finite duration can therefore remain visible after the abundance is fixed. We formulate this resolved problem in a gauge invariant Stueckelberg regime and generate the mass history from a heavy field trajectory. This keeps the field content identical to the vector calculation while giving the time dependent mass a dynamical origin. Transient departures from slow roll occur naturally in inflationary models with localized features, primordial black hole production, and nonattractor phases \cite{Sasaki2018,CarrKuhnel2020,OzsoyTasinato2023,Starobinsky1992,ByrnesColePatil2019,Tasinato2020,JacksonAssadullahiGowKoyamaVenninWands2024}.

Realistic inflationary dynamics generate mass transitions with finite duration. Departures from slow roll arise from temporary changes in inflaton velocity, heavy-field interactions, turns in field space, or localized deformations of the inflationary potential. These processes evolve over a resolved interval. Scalar feature models likewise show that the integrated feature, its duration, and its differentiability control the resulting oscillatory structure \cite{Starobinsky1992,MukhanovFeldmanBrandenberger1992}. Studies of ultra-slow-roll and non-attractor inflation further establish that modes crossing the horizon during an active transition retain information about its temporal evolution \cite{Kinney2005,MartinMotohashiSuyama2013,MotohashiStarobinskyYokoyama2015,PattisonVenninAssadullahiWands2019}.

The finite duration problem has a mode by mode coherence structure. The impulse $\Pi$ measures the zero frequency content of the logarithmic mass excursion and the width $\Delta N$ controls its finite frequency response. Two mass histories can share the same asymptotic masses and the same impulse while producing different responses for modes that resolve the event. Long period modes add coherently and recover the sharp result. Resolved modes acquire phase differences across the transition and their response is filtered by the Fourier transform of the localized mass variation. The differentiability of the background fixes the ultraviolet envelope of this filter. The calculation therefore separates two pieces of information that collapse into one localized source in the sharp limit. The profile transform $\FD$ retains the resolved frequency dependence while $\Pi$ retains the integrated strength. This separation propagates differently into the late observables. The relic abundance samples the weighted one spectrum moment $\AD$ and the induced gravitational wave source samples the two spectrum convolution $\CD$. Imposing the observed dark matter abundance changes the required impulse through $\AD$ and leaves the ratio $\RD=\CD/\AD^2$ in the tensor prediction. Finite duration information can consequently survive abundance normalization. The calculation below derives this residual dependence from the same resolved transition that generates the vector spectrum.

\section{Massive vector dynamics during inflation}
\label{sec:setup}

\subsection{Action and time-dependent mass parametrization}

We consider a massive Abelian vector field evolving as a spectator sector during inflation. The mode calculation below contains the three polarizations of a massive vector. We therefore use the unitary gauge form of a gauge invariant Abelian Stueckelberg theory \cite{RueggRuizAltaba2004}. This maps the field content of the submitted vector calculation to a gauge invariant mass generation mechanism before introducing the time dependence of the mass. The spectator vector dynamics are governed by
\begin{equation}
 S_A=\int\dd^4x\sqrt{-g}\left[-\frac14F_{\mu\nu}F^{\mu\nu}-\frac12M^2(\tau)A_\mu A^\mu\right],
\label{eq:proca_action}
\end{equation}
where the field strength tensor is defined by $F_{\mu\nu}=\partial_\mu A_\nu-\partial_\nu A_\mu$. We take the inflationary background to be a spatially flat Friedmann-Robertson-Walker geometry,
\begin{equation}
 \dd s^2=a^2(\tau)(-\dd\tau^2+\delta_{ij}\dd x^i\dd x^j),
 \qquad \cH\equiv a'/a.
\end{equation}

To separate the late physical mass from its inflationary excursion, we write the vector mass in the form
\begin{equation}
 M^2(\tau)=\frac{m^2J^2(\tau)}{a^2(\tau)},
\label{eq:mass_param}
\end{equation}
where $m$ denotes the asymptotic late time physical mass of the vector field. The ratio $J/a=M/m$ is the physical mass in units of its final value. This form therefore isolates the background mass history from the late dark matter mass and matches the variables used in the rapid mass production mechanism \cite{GrahamMardonRajendran2016,OzsoyTasinato2024,LaRosaTasinato2025}. We choose $J=a$ after the transition so that $M=m$ once the inflationary background returns to adiabatic evolution. The function $J$ will be generated below by the background dependence of the Stueckelberg coefficient.

A concrete gauge invariant completion makes the origin of Eq.~\eqref{eq:mass_param} explicit. Introduce a dimensionless Stueckelberg field $\theta$ and a trigger field $\chi$ through
\begin{equation*}
\begin{aligned}
 \mathcal{L}_{\rm St}={}&-\frac14F_{\mu\nu}F^{\mu\nu}
 -\frac12 f^2(\chi)\bigl(\partial_\mu\theta-g_D A_\mu\bigr)^2\\
 &-\frac12(\partial\chi)^2-U(\chi,\phi),
\end{aligned}
\end{equation*}
which is invariant under $A_\mu\rightarrow A_\mu+\partial_\mu\alpha$ and $\theta\rightarrow\theta+g_D\alpha$. In unitary gauge, $\theta=0$, the vector mass is $M(\chi)=g_D f(\chi)$. Identifying $m=g_D f_+$ and $f(\chi)/f_+=J/a$ reproduces Eq.~\eqref{eq:mass_param} directly. The Stueckelberg phase is the longitudinal degree of freedom. The effective field content then consists of two transverse vector polarizations and one longitudinal vector polarization.

A localized inflationary event can move the minimum of a heavy field over a short interval. This is a standard effective description of heavy field background response during transient multifield evolution \cite{Achucarro2012,PalmaSypsas2020}. We use the local realization
\begin{equation}
 U(\chi,\phi)=V(\phi)+\frac{m_\chi^2}{2}
 \left[\chi-\chi_- -\frac{\Delta\chi}{2}
 \left(1+\tanh\frac{\phi-\phi_1}{\Delta\phi}\right)\right]^2.
\label{eq:trigger_potential}
\end{equation}
The trigger obeys
\begin{equation}
 \ddot\chi+3H_I\dot\chi+m_\chi^2
 \bigl[\chi-\chi_{\min}(\phi)\bigr]=0.
\label{eq:trigger_eom}
\end{equation}
For a monotonic inflaton trajectory through the localized event, $\phi-\phi_1\simeq(\dd\phi/\dd N)_1N$ over the transition interval. The moving minimum in Eq.~\eqref{eq:trigger_potential} then has the profile used below with $\Delta N=\Delta\phi/|(\dd\phi/\dd N)_1|$. The condition $m_\chi\gg H_I/\Delta N$ makes the relaxation time of $\chi$ shorter than the motion of its minimum, so $\chi$ tracks this profile while its fluctuations remain heavy. At the center of the transition $|\dd\chi/\dd N|=|\Delta\chi|/(2\Delta N)$. The kinetic part of the trigger energy therefore satisfies
\begin{equation}
 \frac{\rho_{\chi,\mathrm{kin}}}{3\Mpl^2H_I^2}
 \simeq\frac{\Delta\chi^2}{24\Mpl^2\Delta N^2},
\label{eq:trigger_energy}
\end{equation}
up to tracking corrections suppressed by the heavy mass hierarchy. We require this quantity together with the potential lag energy to remain small compared with unity. Section~\ref{sec:nonslowroll} maps this trajectory to $J/a$. A Higgs completion contains a radial scalar whose evolution can modify the vector spectrum and abundance when that mode is dynamically active \cite{SatoTakahashiYamada2022,RediTesi2022}. The vector only regime considered here is described directly by the Stueckelberg theory. The same effective vector equations follow from a Higgs completion when the radial mass is large compared with both $H_I$ and $H_I/\Delta N$ and its energy density remains subdominant. Throughout the analysis, the vector field has vanishing homogeneous expectation value so that isotropy of the background spacetime remains preserved.

We treat the vector as a spectator sector. This requires that the energy density stored in the produced vector excitations remains subdominant relative to the inflationary background energy density,
\begin{equation}
\rho_A \ll 3\Mpl^2 H_I^2,
\end{equation}
while the resulting late-time abundance remains consistent with the observed dark matter density. These conditions coincide with those underlying the original inflationary vector dark matter framework and guarantee that the backreaction of the vector field on the inflationary background can be consistently neglected \cite{GrahamMardonRajendran2016}. This spectator assumption allows the inflationary background to be treated independently when deriving the vector mode evolution.

\subsection{Longitudinal sector and canonical reduction}

To isolate the physical propagating degrees of freedom, we decompose the spatial vector field into transverse and longitudinal components according to
\begin{equation}
 A_i=A_i^T+\partial_i\varphi,\qquad \partial_iA_i^T=0.
\end{equation}
In the light mass regime relevant for inflationary particle production, transverse vector polarizations remain approximately conformally invariant and experience negligible gravitational production. The longitudinal polarization is physical in the massive phase. In the Stueckelberg completion it is the $\theta$ degree of freedom encoded in $A_\mu$ after unitary gauge fixing. After the nondynamical field $A_0$ is eliminated, the coefficient multiplying the longitudinal kinetic term is proportional to $m^2J^2/(k^2+m^2J^2)$. Since $M^2=m^2J^2/a^2$ remains positive throughout the profile, this coefficient is positive for every momentum and at every time during the transition. The longitudinal quadratic kinetic term therefore keeps the same positive sign across the mass excursion. This mode provides the dominant contribution to inflationary vector dark matter production and shares the constrained structure encountered in previous analyses of massive vector cosmologies \cite{HimmetogluContaldiPeloso2009,GrahamMardonRajendran2016}.

Working in Fourier space, the temporal component $A_0$ is nondynamical and satisfies the algebraic constraint
\begin{equation}
 A_{0,k}=\frac{k^2}{k^2+m^2J^2}\varphi_k'.
\label{eq:A0_constraint}
\end{equation}
Substituting this constraint back into the action yields the effective longitudinal action
\begin{equation}
S_L=\frac12\int\dd\tau\dd^3k\,k^2J^2
\left[\frac{m^2}{k^2+m^2J^2}\varphi_k'\varphi_{-k}'-m^2\varphi_k\varphi_{-k}\right].
\label{eq:long_action_phi}
\end{equation}

To bring the system into canonical form, we introduce the normalized field variable
\begin{equation}
 \pi_k=Z_k\varphi_k,
 \qquad
 Z_k=\frac{kmJ}{\sqrt{k^2+m^2J^2}},
\label{eq:Zdef}
\end{equation}
which transforms the action into the standard quadratic oscillator form
\begin{equation}
 S_L=\frac12\int\dd\tau\dd^3k\left[\pi_k'\pi_{-k}'-\Omega_k^2(\tau)\pi_k\pi_{-k}\right].
\label{eq:long_action_pi}
\end{equation}
The canonical transformation yields
\begin{equation}
 \Omega_k^2=k^2+m^2J^2+\frac{3k^2m^2J'^2}{(k^2+m^2J^2)^2}
 -\frac{k^2}{k^2+m^2J^2}\frac{J''}{J}.
\label{eq:omega_general}
\end{equation}
Appendix~\ref{app:proca} derives Eqs.~\eqref{eq:A0_constraint}-\eqref{eq:omega_general}, including the canonical normalization and the $Z_k''/Z_k$ contribution.

The dominant production regime corresponds to modes for which the vector remains effectively light during inflation,
\begin{equation}
 \frac{mJ}{aH_I}\ll\frac{k}{aH_I}\lesssim1,
\label{eq:light_hierarchy}
\end{equation}
under which Eq.~\eqref{eq:omega_general} simplifies considerably and reduces to
\begin{equation}
 \pi_k''+\left(k^2-\frac{J''}{J}\right)\pi_k=0.
\label{eq:light_mode}
\end{equation}
This equation describes a time-dependent harmonic oscillator whose effective potential is entirely determined by the mass evolution encoded in $J(\tau)$. In the conventional slow-roll limit where $J=a$, one recovers the standard de Sitter relation $J''/J=2/\tau^2$. Departures from slow roll generate an additional localized contribution to this effective potential, thereby modifying the production of longitudinal vector modes.

\section{Finite-duration non-slow-roll mass evolution}\label{sec:nonslowroll}

\subsection{Resolved logarithmic-time transition profile}

We now generalize the instantaneous non-slow-roll construction by allowing the effective vector mass to evolve continuously over a finite interval. The transition is centered at $N=0$, where
\begin{equation}
N=\ln(a/a_1),
\end{equation}
and the characteristic scale is defined through $k_1=a_1H_I$.

The time dependence of the effective mass is parametrized by writing
\begin{equation}
 J(N)=a(N)\,\omega^{1/2}(N),
\label{eq:Jomega}
\end{equation}
such that the late-time adiabatic limit corresponds to $\omega=1$, reproducing the standard physical vector mass. To model a finite-duration departure from slow roll, we introduce
\begin{equation}
 \ln\omega(N)=-2\Pi\left[S_\Delta(N)-1\right],
\label{eq:omega_profile}
\end{equation}
where the transition profile satisfies
$
S_\Delta(-\infty)=0,
\,
S_\Delta(\infty)=1$, with its derivative localized near the transition region.

This parametrization ensures that the post-transition limit satisfies $J/a\rightarrow1$, while the pre-transition regime approaches
$
J/a=\exp(\Pi)$. Since constant offsets do not contribute to derivatives of $J$, the mode equation depends only on the localized transition itself. The quantity
\begin{equation}
 \Pi=-\frac12\int_{-\infty}^{\infty}\dd N\,\frac{\dd\ln\omega}{\dd N}
\label{eq:Pi_def}
\end{equation}
defines the total integrated mass impulse controlling the strength of the non slow roll event. For the monotonic profile used here, $S_\Delta'(N)>0$ and $\dd\ln\omega/\dd N=-2\Pi S_\Delta'(N)$. The decreasing mass branch relevant for the enhancement mechanism therefore has $\Pi>0$, while $\Pi=0$ recovers the slow roll mass history. Because $M/m=J/a=\sqrt{\omega}$, the same parameter is the logarithmic physical mass excursion,
\begin{equation*}
 \Pi=\ln\!\left(\frac{M_{\rm in}}{m}\right).
\end{equation*}
For modes around the transition scale, the light vector condition is $M_{\rm in}=m\exp(\Pi)\ll H_I$. The equality $M_{\rm in}=H_I$ occurs at $\Pi=\ln(H_I/m)$, so the light vector calculation requires the positive branch to remain below this scale with the hierarchy appropriate to the mode expansion. The microscopic realization gives an additional restriction through the trigger spectator condition and through the validity of the background dependent coefficient $f(\chi)$. The parameter $\Pi$ measures the logarithmic excursion of the physical mass. Its microscopic range is therefore fixed jointly by the light vector hierarchy and the trigger relations derived below.

As a minimal smooth realization, we adopt the hyperbolic tangent profile
\begin{equation}
 S_\Delta(N)=\frac12\left[1+\tanh\left(\frac{N}{\Delta N}\right)\right],
\label{eq:intro_profile}
\end{equation}
whose derivative is
\begin{equation}
 S_\Delta'(N)=\frac{1}{2\Delta N}\,\mathrm{sech}^2\left(\frac{N}{\Delta N}\right).
\label{eq:Sprime}
\end{equation}
The instantaneous approximation used in previous analyses is recovered in the limiting case
$
S_\Delta'(N)\rightarrow\delta(N)$, at fixed impulse $\Pi$.

The profile contains two independent pieces of physical information. The parameter $\Pi$ is the zero frequency content of the localized logarithmic mass variation and fixes its integrated strength. The parameter $\Delta N$ fixes the interval over which that variation occurs. In the sharp limit every mode samples the same localized impulse. At finite duration, modes with $x\Delta N\ll1$ remain coherent across the event and recover that limit. Modes with $x\Delta N$ of order unity or larger accumulate phase while the background is evolving and begin to resolve its internal time structure. The profile transform derived below quantifies this loss of coherence. Its large momentum envelope is controlled by the differentiability of the transition.

The Stueckelberg completion above maps the trigger trajectory into the vector mass. Across the finite trigger interval we use a constant logarithmic slope for the positive Stueckelberg scale,
\begin{equation}
 \frac{\partial\ln f}{\partial\chi}=-\frac{\beta}{\Lambda}.
\label{eq:log_slope}
\end{equation}
Integration along the tracked background trajectory gives
\begin{align}
 \frac{f(\chi(N))}{f_+}=\frac{J(N)}{a(N)}
 &=\exp\left[-\frac{\beta}{\Lambda}(\chi(N)-\chi_+)\right],\\
 \chi(N)&=\chi_-+\frac{\Delta\chi}{2}
 \left[1+\tanh\left(\frac{N}{\Delta N}\right)\right].
\label{eq:micro_mapping}
\end{align}
The second line is the tracking solution generated by Eqs.~\eqref{eq:trigger_potential} and \eqref{eq:trigger_eom}. Once the trigger potential and the logarithmic slope in Eq.~\eqref{eq:log_slope} are specified, the time dependence of $J/a$ follows from the field solution. Substitution gives
\begin{equation}
\ln\omega=2\ln(J/a)=-2\Pi\left[S_\Delta-1\right],
\end{equation}
with
\begin{equation}
 \Pi=\frac{\beta\Delta\chi}{\Lambda}
 =-\int_{\chi_-}^{\chi_+}\dd\chi\,\frac{\partial\ln f}{\partial\chi}.
\label{eq:Pi_micro}
\end{equation}
The subtraction of $\chi_+$ enforces $J/a\rightarrow1$ at late times. The trajectory of the heavy field fixes both the logarithmic mass excursion and the duration entering $J(N)$. The mapping also separates the size of the mass hierarchy from the energy stored in the trigger motion. Combining Eqs.~\eqref{eq:trigger_energy} and \eqref{eq:Pi_micro} gives
\begin{equation}
 \frac{\rho_{\chi,\mathrm{kin}}}{3\Mpl^2H_I^2}
 \simeq\frac{\Pi^2}{24\Delta N^2}
 \left(\frac{\Lambda}{\beta\Mpl}\right)^2.
\label{eq:trigger_energy_pi}
\end{equation}
Equation~\eqref{eq:trigger_energy_pi} gives a direct microscopic test of the spectator condition at fixed $\Pi$ and $\Delta N$. The logarithmic mass hierarchy is controlled by $\beta\Delta\chi/\Lambda$, while the background energy carried by the trigger is controlled by $\Delta\chi/(\Mpl\Delta N)$.

Equivalent effective histories can arise when rapid turns in multifield inflation shift a heavy field minimum or when a localized inflaton feature changes a mass controlling coupling. The hyperbolic tangent is the smooth tracking profile generated by the explicit moving minimum in Eq.~\eqref{eq:trigger_potential}. The microscopic theory fixes $\beta$, $\Delta\chi/\Lambda$, $\Delta N$, and the trigger energy. The vector calculation receives this information through $\Pi$ and the resolved transition profile.

\subsection{Mode equation in e-fold time}

In quasi-de Sitter expansion the conformal time satisfies
$a=-1/(H_I\tau)$,
which allows derivatives with respect to conformal time to be rewritten in terms of e-fold time as
\begin{equation}
 \frac{\dd}{\dd\tau}=-\frac{1}{\tau}\frac{\dd}{\dd N},
 \qquad
 \frac{\dd^2}{\dd\tau^2}=\frac{1}{\tau^2}\left(\frac{\dd^2}{\dd N^2}+\frac{\dd}{\dd N}\right).
\label{eq:tauN}
\end{equation}

Writing the time-dependent mass function in exponential form, $J=\exp q(N)$, one obtains the useful identity
\begin{equation}
 \frac{J''}{J}=\frac{q_{NN}+q_N^2+q_N}{\tau^2}.
\label{eq:Jidentity}
\end{equation}

For the finite-duration transition profile, $q(N)=N-\Pi S_\Delta(N)+\Pi$, the effective potential entering the longitudinal mode equation contains contributions proportional to both $S_\Delta'(N)$ and $S_\Delta''(N)$. Terms linear in $\Pi$ generate the leading impulse contribution responsible for the first-order enhancement of the longitudinal mode, while terms quadratic in $\Pi$ determine higher-order corrections that control the full spectral structure.

In the limit $\Delta N\rightarrow0$, these contributions reduce to the instantaneous matching structure previously derived in the non-slow-roll analysis of Ref.~\cite{LaRosaTasinato2025}. For finite $\Delta N$, however, the singular source is replaced by a smooth distribution extending over a finite interval. The resulting mode evolution therefore carries direct information about the temporal structure and differentiability properties of the underlying inflationary transition, providing the central physical mechanism studied throughout the remainder of this work.

\section{Profile transform and finite-width longitudinal spectrum}\label{sec:profile}

\subsection{Transition dynamics as a scattering problem}

The longitudinal mode equation in Eq.~\eqref{eq:light_mode} admits a useful interpretation as an effective one-dimensional scattering problem in conformal time. The time-dependent mass evolution introduces a localized perturbation which mixes the positive- and negative-frequency de Sitter solutions during inflation. To make this structure explicit, we separate the effective potential as
\begin{equation}
 \frac{J''}{J}=\frac{2}{\tau^2}+U_{\rm tr}(\tau),
\end{equation}
where the first term corresponds to the standard de Sitter contribution, while $U_{\rm tr}$ encodes the localized non-adiabatic transition associated with finite-duration mass evolution.

From this perspective, the transition acts as an effective scattering potential. Within an impulse expansion, each localized insertion contributes with a weight determined by the Fourier transform of the transition profile in the time variable that resolves the background evolution. Since the non-adiabatic mass variation is localized in e-fold time $N$, the relevant quantity is the profile transform
\begin{equation}
 \widetilde S_\Delta(\nu)=\int_{-\infty}^{\infty}\dd N\,S_\Delta'(N)\e^{\ii\nu N}.
\label{eq:profile_transform_general}
\end{equation}

The dimensionless frequency variable $\nu$ is naturally determined by the physical modes crossing the transition scale and is approximately given by
\begin{equation}
\nu\simeq\frac{k}{k_1}.
\end{equation}
This identification follows directly from de Sitter kinematics. Defining the standard dimensionless variable
\begin{equation}
y\equiv-k\tau=\left(\frac{k}{k_1}\right)e^{-N}=xe^{-N},
\end{equation}
the oscillatory phase during a localized transition satisfies
\begin{equation}
e^{\ii y}=e^{\ii x}e^{-\ii xN}\left[1+O(xN^2)\right],
\end{equation}
for sufficiently short transitions satisfying $|N|\lesssim\Delta N\ll1$. Consequently, the leading finite-duration correction is governed by the Fourier transform of $S_\Delta'(N)$ evaluated at frequency $\nu\simeq x$, while higher-order corrections are suppressed by terms of order $x\Delta N^2$. The validity of this approximation is tested explicitly through the numerical mode evolution discussed later in Sec.~\ref{sec:numerical}.

For the hyperbolic tangent transition introduced in Eq.~\eqref{eq:Sprime}, the profile transform becomes
\begin{equation}
\begin{split}
 \FD(x)&=\int_{-\infty}^{\infty}\dd N\,\frac{1}{2\Delta N}\mathrm{sech}^2\left(\frac{N}{\Delta N}\right)\e^{\ii xN} \\
 &=\frac{\pi\Delta N x/2}{\sinh(\pi\Delta N x/2)}.
\end{split}
\label{eq:formfactor_main}
\end{equation}

This function acts as a momentum-dependent transfer factor encoding how efficiently a given mode responds to the finite-time background transition. The limiting behaviors follow immediately:
\begin{equation}
 \FD(x)=1-\frac{\pi^2\Delta N^2x^2}{24}+\frac{7\pi^4\Delta N^4x^4}{5760}+O(\Delta N^6x^6),
\label{eq:Fsmall}
\end{equation}
for unresolved long-wavelength modes, while for sufficiently resolved short-wavelength modes one finds
\begin{equation}
 \FD(x)=\pi\Delta N x\,\exp\left[-\frac{\pi\Delta N x}{2}\right]\left[1+O\left(\e^{-\pi\Delta N x}\right)\right].
\label{eq:Flarge}
\end{equation}

The physical interpretation is immediate. Modes satisfying $x\Delta N\ll1$ cannot resolve the internal structure of the transition and therefore experience the same effective impulse as in the instantaneous approximation. In contrast, modes with $x\Delta N\gg1$ resolve the finite duration of the background event and their response is exponentially suppressed. Thus, a smooth transition naturally removes the ultraviolet artifacts associated with singular matching.

\begin{figure}[htbp]
\centering
\includegraphics[width=\linewidth]{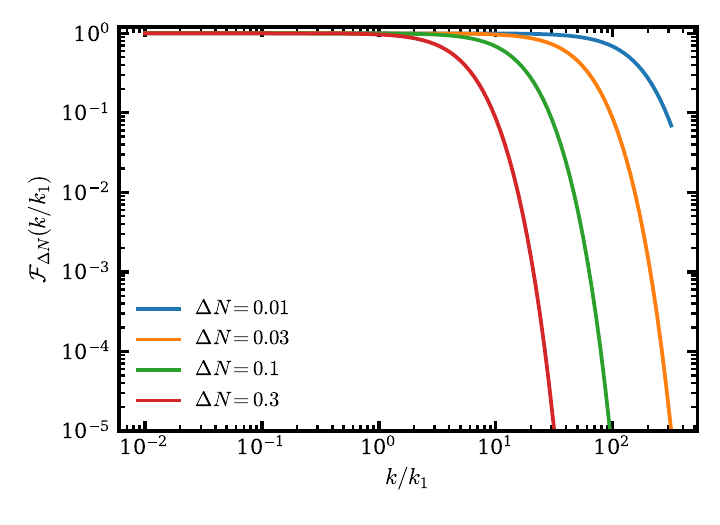}
\caption{Finite-duration transfer kernel associated with a smooth hyperbolic tangent mass transition. The instantaneous approximation is recovered in the unresolved regime $k\Delta N/k_1\ll1$, whereas modes that resolve the internal transition structure are exponentially filtered.}
\label{fig:formfactor}
\end{figure}

\begin{figure}[htbp]
\centering
\includegraphics[width=\linewidth]{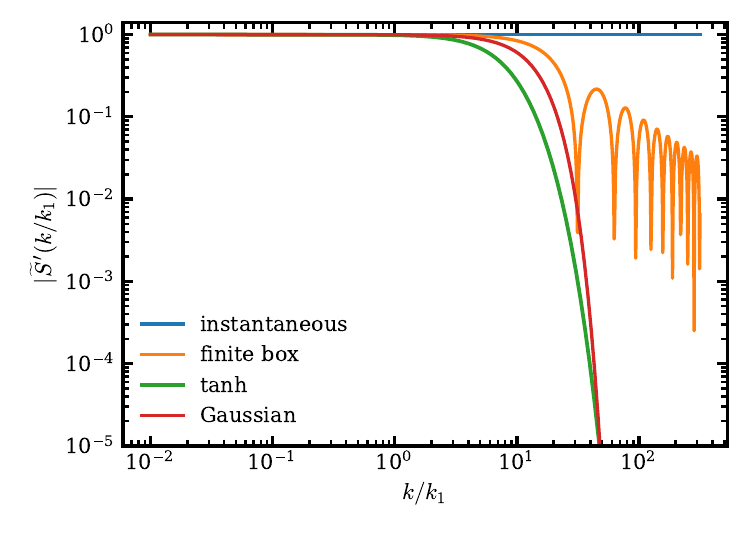}
\caption{Comparison of profile-transform envelopes for transitions sharing the same characteristic width. Instantaneous transitions exhibit no ultraviolet suppression, finite box profiles generate oscillatory power-law behavior, while analytic profiles such as hyperbolic tangent or Gaussian transitions suppress resolved modes exponentially or faster.}
\label{fig:profiles}
\end{figure}

More generally, the asymptotic behavior of the profile transform carries direct information about the differentiability properties of the underlying inflationary background. Analytic transition profiles generate exponential suppression at high momenta, whereas profiles with only finite differentiability produce power-law ultraviolet envelopes. Figure~\ref{fig:formfactor} illustrates the finite-duration kernel associated with the hyperbolic tangent profile used throughout the present analysis, while Fig.~\ref{fig:profiles} compares this behavior with other representative transition profiles of equal width.

\begin{figure}[htbp]
\centering
\includegraphics[width=\linewidth]{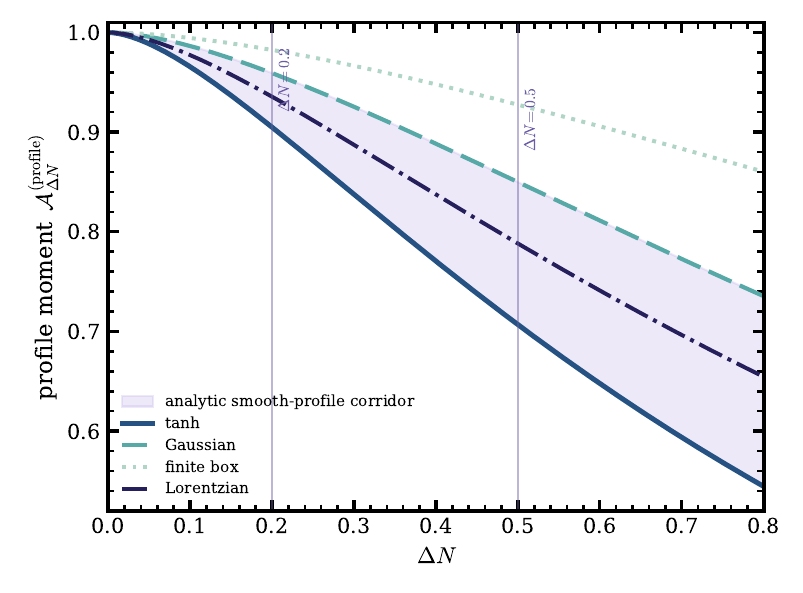}
\caption{Dependence of the abundance moment on the choice of transition profile. The shaded band represents smooth analytic transition families, while the solid curve corresponds to the hyperbolic tangent profile used in the main numerical analysis. The instantaneous limit is recovered for $\Delta N\rightarrow0$ for all normalized smooth profiles.}
\label{fig:profilemoments}
\end{figure}

An important observation is that the low-frequency moment structure is considerably more universal than the detailed ultraviolet behavior. Provided the normalized profile satisfies unit area normalization, all sufficiently smooth transitions obey
\begin{equation}
\widetilde S_\Delta(0)=1,
\end{equation}
which guarantees that the unresolved infrared branch remains unchanged. The ultraviolet regime, however, depends sensitively on the differentiability class of the underlying transition. Discontinuous derivatives generate oscillatory algebraic ringing, whereas analytic background evolution produces exponential suppression. Figure~\ref{fig:profilemoments} shows the corresponding abundance moments for several representative profile families.

The hyperbolic tangent profile adopted throughout this work provides a particularly convenient realization because it yields an exact closed-form transform and furnishes a monotonic analytic interpolation between the sharp-transition limit and fully resolved finite-duration evolution. Although replacing this profile by another smooth transition changes the precise numerical value of $\AD$ at fixed width, the central physical conclusion remains unchanged: finite-duration inflationary dynamics modify particle production through a positive profile-dependent source moment.

\subsection{Finite-duration modification of the longitudinal spectrum}

The longitudinal power spectrum derived in the sharp-transition approximation can be written in the form
\begin{equation}
 \Pg_{\AL}^{(0)}(k)=\frac{H_I^2k^2}{4\pi^2m^2}\Pi_0\left(\frac{k}{k_1}\right),
\label{eq:Psharp_main}
\end{equation}
where the dimensionless spectral function is given by
\begin{equation}
\Pi_0(x)=1-4x\Pi\cos x\,j_1(x)+4x^2\Pi^2j_1^2(x),
\label{eq:intro_sharp}
\end{equation}
with $j_1(x)$ denoting the spherical Bessel function.

Finite-duration effects modify this result because each impulse insertion now carries an additional transfer factor determined by the profile transform. Incorporating these corrections yields the generalized finite-width spectrum
\begin{equation}
 \Pg_{\AL}^{(0)}(k;\Delta N)=\frac{H_I^2k^2}{4\pi^2m^2}\Pi_{\Delta N}\left(\frac{k}{k_1}\right),
\label{eq:Psmooth_main}
\end{equation}
where
\begin{equation}
\begin{split}
\Pi_{\Delta N}(x)=&1-4x\Pi\FD(x)\cos x\,j_1(x)\\
&+4x^2\Pi^2\FD^2(x)j_1^2(x).
\end{split}
\label{eq:intro_smooth}
\end{equation}

This expression immediately reveals three physically important limits. First, the instantaneous approximation is recovered continuously through
\begin{equation}
\Delta N\rightarrow0,
\end{equation}
at fixed momentum $x$. Second, in the unresolved regime
\begin{equation}
x\Delta N\ll1,
\end{equation}
finite-duration corrections first appear only at order $x^2\Delta N^2$, leaving long-wavelength production essentially unchanged. Third, in the resolved regime
\begin{equation}
x\Delta N\gg1,
\end{equation}
the oscillatory ultraviolet structure inherited from singular matching is exponentially suppressed.

To examine the infrared structure explicitly, we use the small-$x$ expansion
\begin{equation}
 j_1(x)=\frac{x}{3}-\frac{x^3}{30}+\frac{x^5}{840}+O(x^7).
\end{equation}
Substituting this together with Eq.~\eqref{eq:Fsmall} gives
\begin{equation}
\begin{split}
 \Pi_{\Delta N}(x)=&1-\frac{4\Pi}{3}x^2\\
&+\left(\frac{4\Pi^2}{9}+\frac{4\Pi}{5}+\frac{\Pi\pi^2\Delta N^2}{18}\right)x^4+O(x^6).
\end{split}
\label{eq:Pi_small}
\end{equation}

An important consequence follows immediately. Since the full dimensional spectrum in Eq.~\eqref{eq:Psmooth_main} carries an overall prefactor proportional to $k^2$, the blue infrared scaling characteristic of inflationary vector dark matter production remains unchanged. Finite-duration effects modify the intermediate enhanced branch and regulate the ultraviolet structure, but they do not generate the scale-invariant large-scale behavior associated with scalar spectator isocurvature perturbations. The infrared protection mechanism therefore survives even in the presence of physically resolved non-slow-roll transitions.

\begin{figure}[htbp]
\centering
\includegraphics[width=\linewidth]{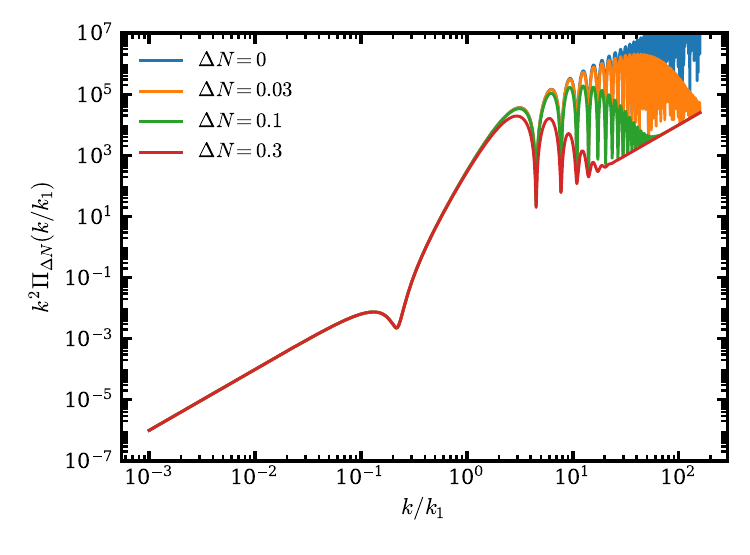}
\caption{Dimensionless longitudinal spectral shape $k^2\Pi_{\Delta N}(k/k_1)$ for impulse strength $\Pi=30$. Finite transition duration preserves the enhanced non-slow-roll amplification while removing the ultraviolet oscillatory structure inherited from the instantaneous approximation.}
\label{fig:spectrum}
\end{figure}

The benchmark $\Pi=30$ used in Fig.~\ref{fig:spectrum} lies on the positive decreasing mass branch. This value places the spectrum on the enhanced ultralight branch while keeping the finite duration filter visible in the same momentum range. It corresponds to $M_{\rm in}/m=\exp(30)=1.07\times10^{13}$. Since $\omega=(J/a)^2=M^2/m^2$, the change of $\omega$ across the transition is $\exp(-60)$ and the physical mass ratio $J/a=M/m$ changes by $\exp(-30)$. At $m=10^{-18}\,{\rm eV}$ and $H_I=10^{12}\,{\rm GeV}$, which give the smallest hierarchy $H_I/m$ among the displayed benchmark ranges, $M_{\rm in}=1.07\times10^{-5}\,{\rm eV}$ and $M_{\rm in}/H_I=1.07\times10^{-26}$. The equality scale $M_{\rm in}=H_I$ occurs at $\Pi=\ln(H_I/m)\simeq89.8$ for the same values, and every smaller displayed mass or larger displayed Hubble scale increases this separation. The benchmark is therefore deep inside the light vector regime used in the mode calculation. Its microscopic interpretation follows from Eqs.~\eqref{eq:Pi_micro} and \eqref{eq:trigger_energy_pi}. The condition $\Pi=30$ fixes the logarithmic excursion $\beta\Delta\chi/\Lambda=30$. For the representative width $\Delta N=0.2$, Eq.~\eqref{eq:trigger_energy_pi} gives
\begin{equation*}
 \frac{\rho_{\chi,\mathrm{kin}}}{3\Mpl^2H_I^2}
 \simeq9.38\times10^2
 \left(\frac{\Lambda}{\beta\Mpl}\right)^2.
\end{equation*}
The spectator requirement therefore constrains the microscopic logarithmic slope independently of the light vector hierarchy. The factor $\exp(-60)$ is the change of the squared mass ratio $\omega$. The benchmark consistency conditions are the light vector hierarchy, the trigger spectator condition, and the validity of the background dependent Stueckelberg coefficient. The ultraviolet assessment of $\beta$ and $\Lambda$ depends on the microscopic realization of that coefficient.

\section{Relic abundance}\label{sec:abundance}

\subsection{Late-time energy density and radiation-era transfer functions}

Following inflation and reheating, the produced longitudinal vector modes evolve through the radiation-dominated epoch before eventually behaving as nonrelativistic dark matter. The present-day relic abundance is therefore determined not only by the inflationary production spectrum derived previously, but also by the subsequent cosmological transfer of these modes as the Universe expands and the vector mass becomes dynamically relevant.

The late-time energy density stored in the longitudinal sector can be written as
\begin{equation}
\rho_{\AL}=\frac{m^2}{2a^2}\int\dd\ln k\left[\frac{\Pg_{\partial_\tau\AL}(k)}{k^2+a^2m^2}+\Pg_{\AL}(k)\right].
\label{eq:rho_general_main}
\end{equation}
This expression incorporates both the field amplitude contribution and the kinetic contribution associated with the longitudinal degree of freedom. Radiation-era transfer functions evolve the inflationary spectrum in Eq.~\eqref{eq:Psmooth_main} into the final relic abundance.

The physically important epoch occurs when the vector mass becomes dynamically comparable to the Hubble expansion rate, namely
\begin{equation}
 H_*=m,\qquad k_*=a_*m,
\label{eq:kstar_def}
\end{equation}
where the star denotes the radiation-era time at which coherent massive oscillations begin. The scale $k_*$ therefore defines the momentum region that dominates the transition between relativistic evolution and nonrelativistic dark matter behavior.

To characterize the relative position of the inflationary non-slow-roll feature with respect to this radiation-era turnover scale, we introduce the dimensionless ratio
\begin{equation}
 \sigma\equiv\frac{k_*}{k_1}=\frac{a_*m}{a_1H_I}.
\label{eq:sigma_def}
\end{equation}
Assuming instantaneous reheating followed by standard radiation domination, the scale factor evolves according to
\begin{equation}
 \frac{a_*}{a_R}=\left(\frac{H_I}{m}\right)^{1/2},\qquad
 \sigma=e^{N_1}\left(\frac{m}{H_I}\right)^{1/2},
\label{eq:sigma_relation}
\end{equation}
where
\begin{equation}
N_1=\ln(a_R/a_1).
\end{equation}

The regime of particular interest corresponds to large values of $\sigma$, where the non-slow-roll enhancement occurs in the momentum region that subsequently dominates the relic-density integral. In the original inflationary vector dark matter framework, the abundance factorizes into a simple function of the vector mass, inflationary Hubble scale, and an integrated spectral contribution associated with the longitudinal peak \cite{GrahamMardonRajendran2016}. Related abundance calculations have also appeared in studies of inflationary dark photons and coherent vector production mechanisms \cite{DrorHarigaya2019,NakaiNambaWang2020,SalehianGorjiFirouzjahiMukohyama2021}. In the sharp-transition approximation, the non-slow-roll abundance can be expressed as
\begin{equation}
\begin{split}
 \rDM&\equiv\frac{\rho_{\AL}}{\rho_{\rm DM}} \\
 &\simeq 10^{-7}\left(\frac12+\Pi+\Pi^2\right)\nonumber\\
&\quad\times\left(\frac{m}{1.5\times10^{-21}{\rm eV}}\right)^{1/2}
\left(\frac{H_I}{10^{14}{\rm GeV}}\right)^2,
\end{split}
\label{eq:sharp_abundance_main}
\end{equation}
which provides the standard large-$\sigma$ limit obtained from fitting the post-inflationary transfer functions.

The central modification introduced by finite-duration transitions arises because the dominant non-slow-roll enhancement of the abundance scales quadratically with the impulse amplitude. As a consequence, the finite-width correction enters through a positive profile-dependent moment of the transfer factor $\FD^2(x)$. We therefore define the dimensionless source moment
\begin{equation}
\AD=\frac{\int_0^\infty\dd\ln x\,W(x)\FD^2(x)}{\int_0^\infty\dd\ln x\,W(x)},
\qquad W(x)=\frac{x^2}{(1+x^2)^2}.
\label{eq:AD_main}
\end{equation}

The weight function $W(x)$ provides an analytic representation of the broad radiation-era momentum contribution centered around the vector turnover scale. Physically, $\AD$ quantifies how efficiently the finite-time transition transfers power into those momentum modes which later dominate the relic abundance. Similar questions concerning abundance renormalization appear whenever a coherently evolving vector sector experiences a post-inflationary non-adiabatic transition, where the matching process itself becomes part of the relic-density calculation \cite{Khan:2026nsz}.

In the full radiation-era evolution, the exact late-time energy density may be written in the form
\begin{equation}
\rho_{\AL}=\frac12\frac{a_*^3}{a^3}\left(\frac{mH_I}{2\pi}\right)^2 I_\rho,
\label{eq:rho_Irho}
\end{equation}
where the dimensionless integral $I_\rho$ incorporates transfer-function contributions from both momentum regions
\begin{equation}
k<k_*,
\qquad
k>k_*.
\end{equation}
In the sharp-transition limit and for large $\sigma$, this integral reduces to
\begin{equation}
 I_\rho\simeq\frac32\left(1+2\Pi+2\Pi^2\right),
\label{eq:Irho_large_sigma}
\end{equation}
up to the normalization convention adopted for the impulse parameter.

Once finite-duration effects are included, the impulse-dominated contribution to this transfer integral becomes weighted by the profile-dependent moment defined in Eq.~\eqref{eq:AD_main}. Replacing the approximate analytic weight $W(x)$ by the exact numerical radiation-era transfer function modifies the numerical value of the curve but leaves the fundamental factorization structure unchanged. In particular, finite-time corrections always enter through the universal source factor $\FD^2(x)$ provided the non-slow-roll event occurs before subsequent cosmological transfer evolution begins.

More general thermal histories modify the transfer-function weight. For example, prolonged reheating or a non-standard equation of state modifies the transfer-function weight appearing in the relic-density integral. However, the finite-time transition remains factorized through the same profile-dependent source factor, implying that the inflationary production mechanism and the subsequent cosmological transfer evolution remain physically separable.

The resulting finite-duration abundance relation becomes
\begin{equation}
\begin{split}
 \rDM&\simeq 10^{-7}\left(\frac12+\Pi+\Pi^2\AD\right)\nonumber\\
&\quad\times\left(\frac{m}{1.5\times10^{-21}{\rm eV}}\right)^{1/2}
\left(\frac{H_I}{10^{14}{\rm GeV}}\right)^2.
\end{split}
\label{eq:smooth_abundance_main}
\end{equation}

This expression clearly separates three physically independent ingredients governing vector dark matter production. The first is the vector mass $m$, the second is the inflationary Hubble scale $H_I$, both already present in the conventional inflationary vector dark matter scenario, while the third is the finite-time transition profile encoded in $\AD$. The latter represents a genuinely new physical parameter absent from instantaneous matching calculations and determines how strongly finite-duration effects modify the required inflationary impulse.

\subsection{Required impulse for fixed relic abundance}

For a fixed target dark matter fraction $\rDM$, Eq.~\eqref{eq:smooth_abundance_main} determines the total impulse strength required to generate the observed abundance. It is convenient to define the dimensionless combination
\begin{equation}
Y\equiv\frac{\rDM}{10^{-7}}
\left(\frac{1.5\times10^{-21}{\rm eV}}{m}\right)^{1/2}
\left(\frac{10^{14}{\rm GeV}}{H_I}\right)^2 .
\label{eq:Ydef}
\end{equation}
The abundance condition then reduces to the quadratic equation
\begin{equation}
 \AD\Pi^2+\Pi+\frac12-Y=0.
\end{equation}
The physically relevant positive solution is
\begin{equation}
 \Pi_{\rm DM}=\frac{\sqrt{1-2\AD+4\AD Y}-1}{2\AD}.
\label{eq:PiDM}
\end{equation}

This relation directly quantifies how finite-duration effects renormalize the required non-slow-roll event. In the phenomenologically relevant large-enhancement regime,
$Y\gg1$, the solution simplifies to
$\Pi_{\rm DM}\simeq\sqrt{\frac{Y}{\AD}}$.
The physical interpretation is particularly transparent. Since a smoother transition suppresses efficient mode production, a larger integrated impulse is required to compensate and reproduce the same relic abundance. However, because the abundance depends quadratically on the impulse, this enhancement enters only through the relatively mild scaling proportional to $\AD^{-1/2}$.

\begin{figure}[htbp]
\centering
\includegraphics[width=\linewidth]{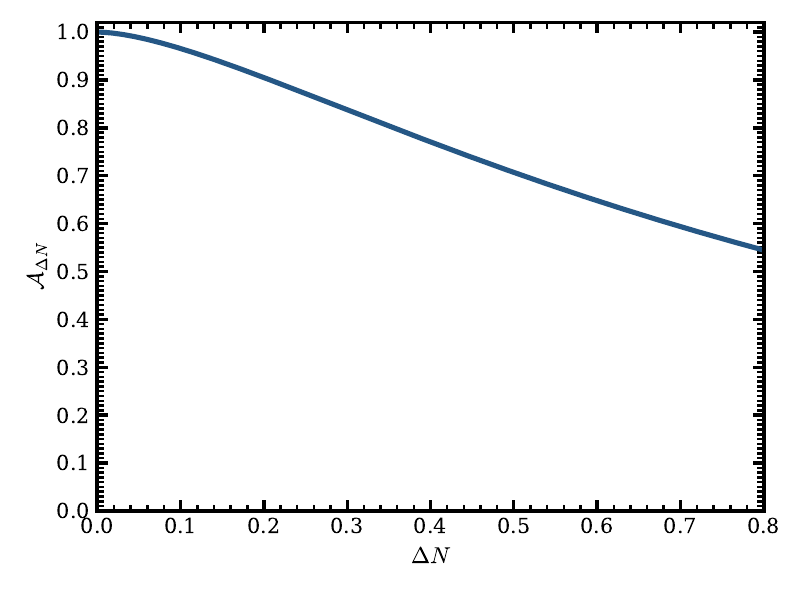}
\caption{Profile-dependent abundance moment $\AD$ entering the relic-density integral, computed from Eq.~\eqref{eq:AD_main}. The sharp-transition normalization is recovered for $\Delta N\ll1$, while increasing transition duration produces a smooth monotonic suppression of the abundance contribution.}
\label{fig:AD}
\end{figure}

\begin{figure}[htbp]
\centering
\includegraphics[width=\linewidth]{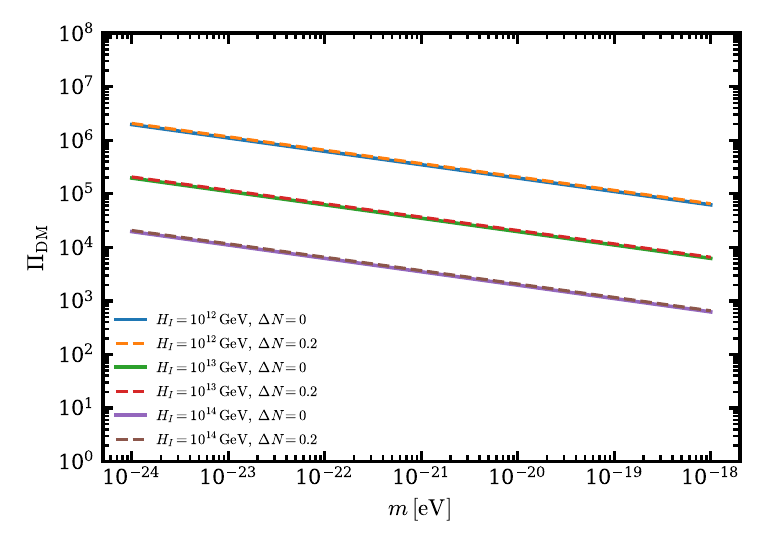}
\caption{Impulse strength required for the longitudinal vector sector to account for the full observed dark matter abundance, obtained from Eq.~\eqref{eq:PiDM}. Solid curves correspond to the sharp-transition approximation, while dashed curves include finite-duration effects with $\Delta N=0.2$.}
\label{fig:PiReq}
\end{figure}

Figure~\ref{fig:AD} shows the direct numerical evaluation of the source moment $\AD$. Figure~\ref{fig:PiReq} illustrates the corresponding impulse strength required for the longitudinal vector sector to account for the entire dark matter abundance across the ultralight mass window. The results demonstrate that finite-duration effects and lower inflationary scales do not qualitatively alter the standard abundance scaling relations. Instead, they shift the magnitude of the non-slow-roll event required to produce a fixed relic fraction.

\begin{figure}[htbp]
\centering
\includegraphics[width=\linewidth]{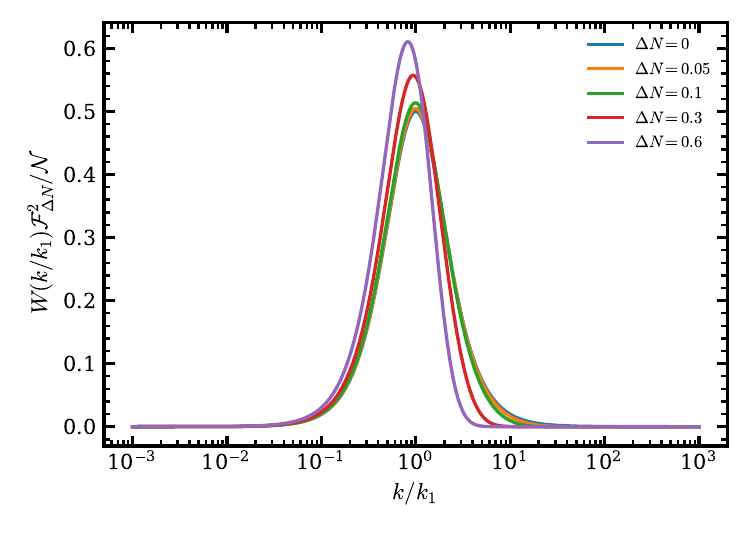}
\caption{Momentum-space contribution to the abundance integral. Finite transition duration shifts the dominant support away from modes that fully resolve the transition, thereby reducing sensitivity of the relic-density calculation to the ultraviolet structure present in the sharp-transition spectrum.}
\label{fig:contrib}
\end{figure}

\section{Induced gravitational waves}\label{sec:gw}

\subsection{Tensor source and sharp-limit normalization}

The enhanced longitudinal vector fluctuations generated during inflation do not only contribute to the relic abundance, but also source tensor perturbations at second order through their anisotropic stress during the radiation-dominated epoch. As a result, the same mechanism responsible for dark matter production leaves an additional gravitational-wave signature which provides an independent probe of the inflationary production history.

During radiation domination, the tensor perturbations satisfy the standard evolution equation
\begin{equation}
 h_\lambda''+2\cH h_\lambda'+k^2h_\lambda=4\Lambda_{ij}^{\lambda}(\bm k)S_{ij}(\tau,\bm k),
\label{eq:tensor_eq}
\end{equation}
where $\Lambda_{ij}^{\lambda}$ denotes the transverse-traceless projection operator associated with polarization state $\lambda$, while $S_{ij}$ represents the anisotropic-stress source generated by quadratic combinations of the vector perturbations.

The general formalism governing induced gravitational-wave production from second-order cosmological perturbations is well established \cite{AnandaClarksonWands2007,Baumann2007,KohriTerada2018,Domenech2021}. The specific radiation-era transfer functions relevant for the longitudinal massive-vector source were derived in Ref.~\cite{MarriottBestPelosoTasinato2025}, while the non-slow-roll vector dark matter scenario of Ref.~\cite{LaRosaTasinato2025} obtained the following phenomenological fit for the induced tensor spectrum,
\begin{equation}
\begin{split}
 \Omega_{\rm GW}(x_*)&\simeq 1.2\times10^{-24}\left(\frac12+\Pi+\Pi^2\right)^2
\left(\frac{H_I}{10^{14}{\rm GeV}}\right)^4 \\
&\quad \times\frac{x_*^{2.6}}{(1+x_*^2)^{1.325}}
\left(1+\frac{x_*^3}{7.29\times10^5}\right)^{-0.65},
\end{split}
\label{eq:GW_fit_sharp_main}
\end{equation}
where the dimensionless frequency variable is defined through $x_*=\frac{f}{f_*}$, with characteristic frequency
\begin{equation}
 f_*\simeq4.7\times10^{-14}
\left(\frac{m}{1.5\times10^{-21}{\rm eV}}\right)^{1/2}{\rm Hz}.
\label{eq:fstar_main}
\end{equation}

This characteristic scale is determined by the epoch at which the vector field becomes nonrelativistic and therefore directly connects the gravitational-wave signal to the underlying vector mass scale.

Finite-duration effects modify the tensor calculation in a manner qualitatively different from the relic abundance calculation. At fixed impulse strength, each vector spectrum entering the anisotropic-stress source is filtered independently according to
\begin{equation}
 \Pg_A(q)\;\longrightarrow\;\Pg_A(q)\FD^2(q/k_1).
\label{eq:tensor_leg_filter}
\end{equation}

Since the tensor source is quadratic in the vector power spectrum, the full radiation era tensor signal depends on a convolution involving two independently filtered spectra. The fixed impulse gravitational wave amplitude therefore takes the schematic form
\begin{equation}
 \Omega_{\rm GW}^{\rm max}(\Pi,\Delta N)
 \simeq \Omega_{\rm GW}^{\rm max}(\Pi,0)\,\CD,
\label{eq:fixed_impulse_tensor}
\end{equation}
where $\CD$ denotes the tensor-source suppression factor associated with the double convolution.

This differs fundamentally from the relic abundance calculation, where finite-duration effects enter only through the single moment $\AD$. Eliminating the impulse parameter using the fixed observed dark matter abundance yields the more physically relevant relation
\begin{equation}
 \Omega_{\rm GW}^{\rm max}\simeq1.2\times10^{-10}\,\rDM^2
 \left(\frac{1.5\times10^{-21}{\rm eV}}{m}\right)\RD,
\label{eq:master_relation_main}
\end{equation}
where the residual finite-width factor is defined by
\begin{equation}
 \RD\equiv\frac{\CD}{\AD^2}.
\label{eq:R_def_main}
\end{equation}

Equations~\eqref{eq:master_relation_main} and \eqref{eq:R_def_main} reveal an important physical point. A simple abundance rescaling would account only for the suppression of particle production due to finite-duration effects. However, once the relic abundance is held fixed, this suppression is largely compensated by the larger required impulse $\Pi_{\rm DM}$ obtained in Eq.~\eqref{eq:PiDM}. The remaining tensor suppression therefore originates from an independent physical effect: the reshaping of the tensor-source convolution itself. This residual dependence, encoded in $\RD$, represents one of the central physical consequences of finite-duration inflationary transitions.

\subsection{Finite-width tensor-source kernels}

The full radiation-era tensor source involves a nontrivial combination of vector transfer functions, angular transverse-traceless projection factors, and Green-function integrations over the source history. Consequently, the finite-duration correction requires a convolution of two independently filtered vector spectra.

To quantify this effect, we first introduce a positive-definite logarithmic kernel defined by
\begin{align}
\mathcal{C}_{\Delta N}(\eta)&=\frac{\mathcal{N}_{\Delta N}(\eta)}{\mathcal{D}(\eta)},\label{eq:Cdiag}\\
\mathcal{N}_{\Delta N}&=\int\dd\ln u\dd\ln v\,K_\eta(u,v)\FD^2(u)\FD^2(v),\nonumber\\
\mathcal{D}&=\int\dd\ln u\dd\ln v\,K_\eta(u,v),\nonumber\\
K_\eta(u,v)&=W(u)W(v)\exp[-\{\ln(u/v)\}^2/(2\eta^2)].\nonumber
\end{align}

This construction provides a controlled diagnostic of the breakdown of exact factorization. For sufficiently broad kernels, one recovers the approximate relation
\begin{equation}
\CD\simeq\AD^2,
\end{equation}
whereas narrow kernels preferentially weight pairs of comparable momentum modes and therefore amplify deviations from simple factorization.

As a second and more realistic numerical test, we retain the full two-dimensional transverse-traceless geometry entering the gravitational-wave source. Introducing the standard momentum variables
\begin{equation}
v=\tilde k/k,
\qquad
u=\frac{|\bm k-\tilde{\bm k}|}{k},
\qquad
\mu=\hat{\bm k}\cdot\hat{\tilde{\bm k}},
\end{equation}
the geometric relation becomes
\begin{equation}
 u=(1+v^2-2\mu v)^{1/2}.
\end{equation}

The corresponding finite-width calibration factor is defined as
\begin{equation}
 \mathcal{C}^{\rm TT}_{\Delta N}(x_*)=
 \frac{\mathcal{I}^{\rm TT}_{\Delta N}(x_*)}
 {\mathcal{I}^{\rm TT}_{0}(x_*)},
\label{eq:TTcalibration}
\end{equation}
where
\begin{align}
\mathcal{I}^{\rm TT}_{\Delta N}(x_*)=&
\int_0^\infty \dd\ln v\int_{-1}^{1}\dd\mu\,
\mathcal{G}(u,v,\mu)\nonumber\\
&\times P_{\rm RD}(ux_*)P_{\rm RD}(vx_*)
\FD^2(ux_*)\FD^2(vx_*),\label{eq:ITT}\\
\mathcal{I}^{\rm TT}_{0}(x_*)=&
\int_0^\infty \dd\ln v\int_{-1}^{1}\dd\mu\,
\mathcal{G}(u,v,\mu)\nonumber\\
&\times P_{\rm RD}(ux_*)P_{\rm RD}(vx_*).
\end{align}

The geometric factor
\begin{equation}
 \mathcal{G}(u,v,\mu)=\frac{v^3(1-\mu^2)^2}{u^3}
\label{eq:TTgeom}
\end{equation}
originates directly from the transverse-traceless projection of the vector anisotropic-stress tensor, while $P_{\rm RD}$ denotes the analytic radiation-era spectral fit derived in Ref.~\cite{MarriottBestPelosoTasinato2025}.

Equation~\eqref{eq:TTcalibration} preserves the full angular structure of the physical tensor source while consistently incorporating finite-duration filtering on both vector legs. By employing the published radiation-era spectral fit, the calculation isolates precisely the source normalization relevant for determining how finite-duration effects propagate into the final gravitational-wave amplitude.

The fixed-abundance residual ratio then becomes
\begin{equation}
 \RD^{\rm TT}(x_*)=\frac{\mathcal{C}^{\rm TT}_{\Delta N}(x_*)}{\AD^2}.
\label{eq:RTT}
\end{equation}

For the representative width parameter $\Delta N=0.2$, the frequency-averaged transverse-traceless source integration gives $\mathcal{R}^{\rm TT}_{\Delta N}\simeq0.21$, whereas the broad positive kernel approximation yields
$\RD\simeq1.04$. Taken together, these calculations define the physically allowed source-kernel interval used in the numerical analysis below and demonstrate explicitly that finite-duration effects continue to modify the tensor source even after fixing the relic abundance.

\begin{figure}[htbp]
\centering
\includegraphics[width=\linewidth]{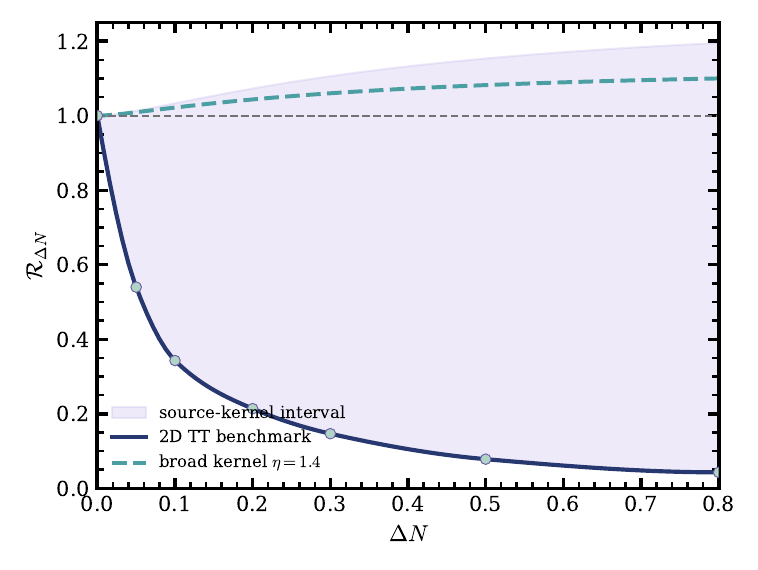}
\caption{Residual tensor-shape factor $\RD=\CD/\AD^2$ evaluated after imposing fixed relic abundance. The shaded band indicates the source-level interval bounded by the broad positive kernel and the full transverse-traceless numerical calibration. The points denote the discrete transverse-traceless calibration values used in normalizing the finite-width gravitational-wave predictions.}
\label{fig:tensorcheck}
\end{figure}

\subsection{Frequency-domain interpretation and observational implications}

Equation~\eqref{eq:master_relation_main} provides the source-level finite-duration amplitude relation governing the induced gravitational-wave signal. This result demonstrates that finite-duration inflationary dynamics modify the gravitational-wave prediction through an independent source-shape observable that survives even after imposing the observed dark matter abundance, thereby establishing a new physical link between inflationary transition dynamics and late-time gravitational-wave observables. The vector mass determines the characteristic frequency scale through Eq.~\eqref{eq:fstar_main}, while the transition duration modifies both the overall amplitude and the ultraviolet structure once both vector legs become capable of resolving the internal transition profile.

The frequency-domain analysis therefore keeps the residual tensor-shape factor explicit and isolates the portion of the source dependence that survives after fixing the relic abundance.

\begin{figure}[htbp]
\centering
\includegraphics[width=\linewidth]{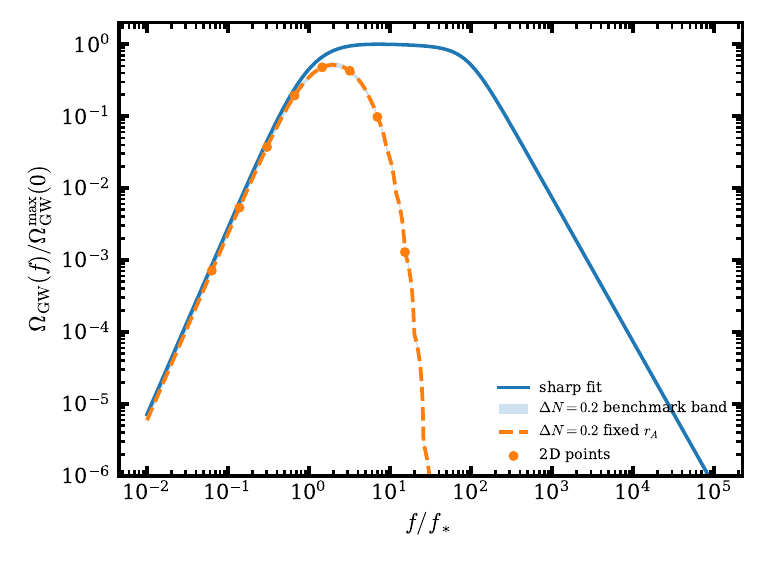}
\caption{Induced gravitational-wave spectrum as a function of $f/f_*$ normalized relative to the sharp-transition $r_A=1$ amplitude. The dashed curve and points correspond to the $\Delta N=0.2$ transverse-traceless source integration, while the shaded region denotes the finite-width source-kernel uncertainty interval.}
\label{fig:gwshape}
\end{figure}

Figure~\ref{fig:gwshape} compares the sharp-transition result with the full $\Delta N=0.2$ transverse-traceless source integration. The sharp limit is retained as the published radiation-era reference normalization, while the finite-duration calculation demonstrates the high-frequency smoothing induced by the filtered tensor source.

\begin{figure}[htbp]
\centering
\includegraphics[width=9cm, height=8cm]{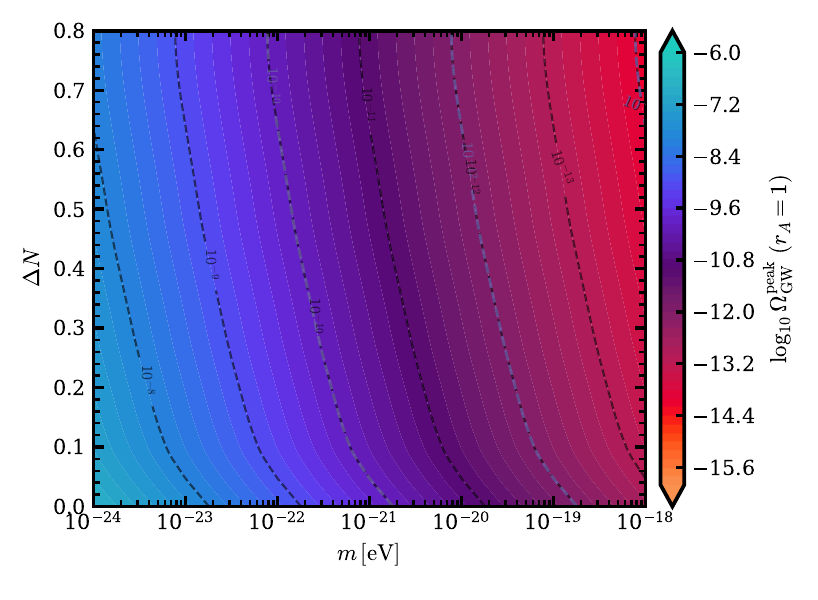}
\caption{Peak gravitational-wave energy density after imposing relic-density saturation for the vector dark matter component, corresponding to $r_A=1$. The vector mass determines the characteristic frequency through Eq.~\eqref{eq:fstar_main}, while finite-duration effects enter through the residual source-shape factor $\RD$.}
\label{fig:GWmap}
\end{figure}

Figure~\ref{fig:GWmap} shows the predicted gravitational-wave amplitude after imposing a relic-density-saturating vector abundance, corresponding to the case $r_A=1$, using the transverse-traceless calculation as the central finite-width normalization.

\begin{figure}[htbp]
\centering
\includegraphics[width=\linewidth]{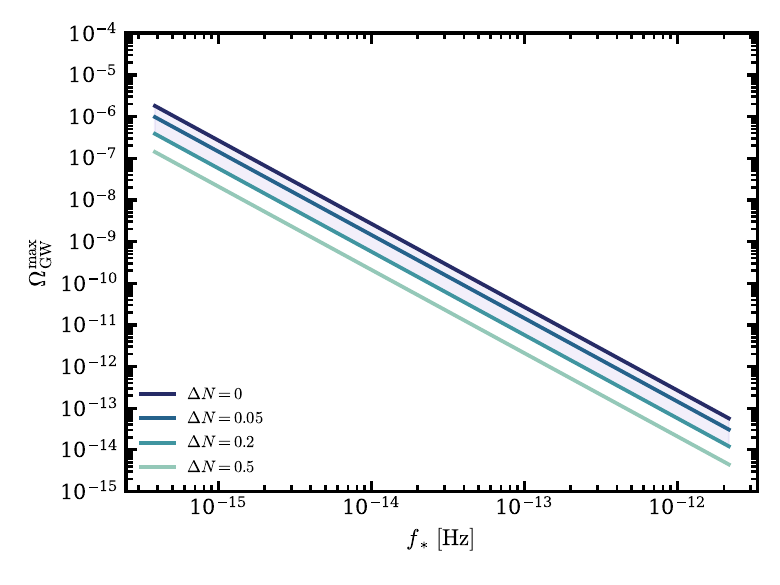}
\caption{Frequency-amplitude relation after imposing the relic-density constraint $r_A=1$. The shaded interval denotes the finite-width source-kernel range for $\Delta N=0.2$, bounded by the full transverse-traceless integration and the broad positive-kernel approximation.}
\label{fig:freqamp}
\end{figure}

Figure~\ref{fig:freqamp} presents the corresponding frequency-amplitude relation together with the kernel-dependent uncertainty band for the representative transition width $\Delta N=0.2$.

These results highlight the complementarity between late-time and early-Universe probes of ultralight vector dark matter. Pulsar-timing measurements of coherent ultralight vector dark matter constrain the local field configuration and polarization properties of the vector sector, whereas the induced gravitational-wave background probes the inflationary production history responsible for generating that sector in the first place \cite{DrorWei2025,NomuraOmiyaTanaka2025}.

\section{Numerical implementation}\label{sec:numerical}

We test the approximations introduced above through seven complementary numerical calculations. Together they assess the profile-transform formalism, the relic-abundance calculation, the tensor-source convolution, and the finite-width corrections used in the analytic treatment.

First, the profile transform is evaluated numerically and compared with its analytic asymptotic limits. Second, the abundance moment defined in Eq.~\eqref{eq:AD_main} is computed through logarithmic quadrature over the support of the radiation-era transfer function. Third, the relic-density-saturating impulse parameter obtained from Eq.~\eqref{eq:PiDM} is evaluated throughout the ultralight vector mass range. Fourth, the finite-width tensor relation is computed using both the broad positive kernel of Eq.~\eqref{eq:Cdiag} and the full transverse-traceless source integration of Eq.~\eqref{eq:TTcalibration}. Fifth, the longitudinal mode equation is solved numerically in logarithmic time in order to compare the exact mode evolution with the analytic envelope prediction $\FD^2$. Sixth, a second mode integration is performed while retaining the nonlinear contribution proportional to $q_N^2$ in order to probe the finite-impulse regime. Finally, the tensor calculation is repeated using an explicit radiation-era retarded Green-function integration to test the source-level approximation employed in the semi-analytic treatment.

The numerical quadrature for the abundance moment yields

\begin{equation}
\begin{array}{c|cccccc}
\Delta N&0&0.05&0.10&0.20&0.30&0.50\\ \hline
\AD&1.000&0.989&0.966&0.905&0.838&0.707\\
\AD^2&1.000&0.977&0.932&0.819&0.702&0.500
\end{array}
\label{eq:AD_table}
\end{equation}

These results quantify the impact of finite duration effects on the production process. We use $\Delta N=0.2$ as a representative benchmark in the resolved transition regime. The event lasts one fifth of an e fold, so modes around the transition scale reach $x\Delta N$ values for which the profile response begins to depart from the coherent limit while the background event remains brief compared with the surrounding inflationary evolution. For this width, the fixed impulse relic abundance is reduced by approximately $10\%$. The tensor source receives an additional suppression whose magnitude depends on the tensor kernel entering the convolution integral.

Once the relic abundance is fixed, the required impulse scales approximately as $\AD^{-1/2}$. The final tensor amplitude then retains the residual ratio $\RD$ introduced previously in addition to the abundance normalization. Numerically, the broad positive kernel approximation gives $\RD=1.044$ for $\Delta N=0.2$, whereas the full transverse traceless source integration yields $\RD\simeq0.21$ after averaging over the gravitational wave peak. The comparison shows that relic density saturating gravitational wave predictions depend on the full source convolution together with the abundance moment.

To test the analytic mode-envelope approximation directly, the longitudinal mode equation is solved numerically in logarithmic time using

\begin{equation}
 \pi_{k,NN}+\pi_{k,N}+
 \left[x^2e^{-2N}-\left(q_{NN}+q_N^2+q_N\right)\right]\pi_k=0,
\label{eq:mode_N_numeric}
\end{equation}

where $x=k/k_1$ and the background profile is parameterized by

\begin{equation}
q(N)=N-\Pi S_\Delta(N)+\Pi .
\end{equation}

Bunch-Davies initial conditions are imposed in the deep subhorizon regime where $xe^{-N}\gg1$. The late-time mode power is then compared against a narrow-transition reference solution. Within the linear-response regime, we define the normalized envelope ratio

\begin{equation}
 \mathcal{E}_{\rm num}(x,\Delta N)=
 \frac{|\pi_k(\Delta N)|^2-|\pi_k(\Pi=0)|^2}
 {|\pi_k(\Delta N_{\rm ref})|^2-|\pi_k(\Pi=0)|^2}
\label{eq:mode_envelope_ratio}
\end{equation}

which provides a direct numerical test of the analytic finite-width prediction.

The numerical solutions confirm that $\mathcal{E}_{\rm num}(x,\Delta N)$ accurately reproduces the analytic envelope $\FD^2(x)$ throughout both the unresolved and transition regimes. The median absolute deviations across the sampled momentum grid are found to be $1.1\%$, $2.0\%$, and $2.6\%$ for transition widths $\Delta N=0.1$, $0.2$, and $0.4$, respectively. The largest deviations occur in the strongly resolved ultraviolet region where the physical response is already exponentially suppressed.

\begin{figure}[htbp]
\centering
\includegraphics[width=\linewidth]{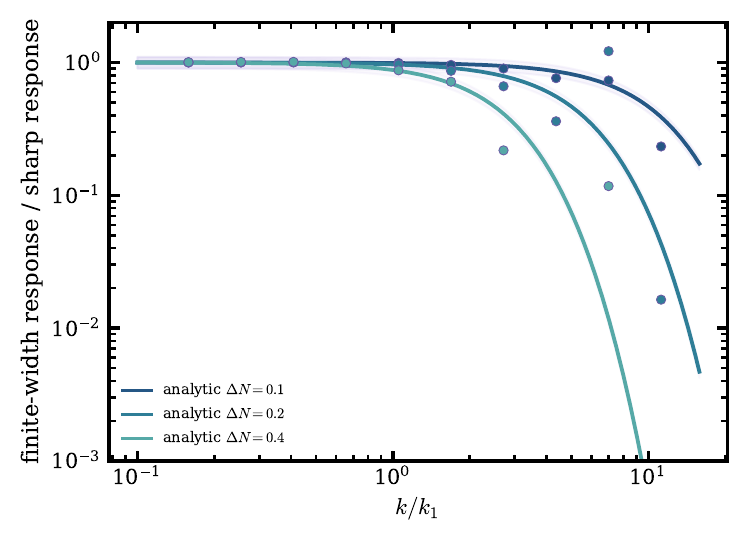}
\caption{Numerical validation of the finite-width envelope approximation. Points correspond to direct numerical solutions of Eq.~\eqref{eq:mode_N_numeric} in the small-impulse regime normalized against a narrow-transition reference solution. Solid curves denote the analytic prediction $\FD^2$, while the shaded region indicates ten percent deviations.}
\label{fig:modecheck}
\end{figure}

\begin{figure}[htbp]
\centering
\includegraphics[width=10cm, height=8cm]{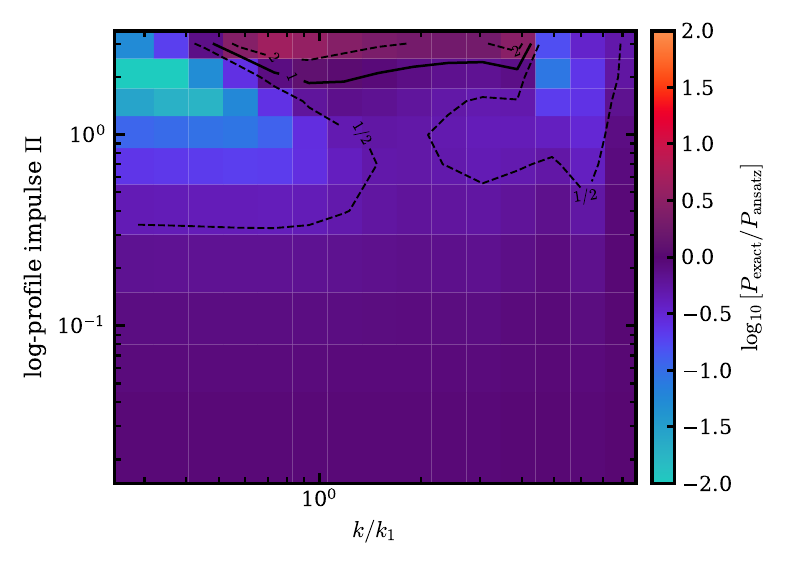}
\caption{Finite-impulse mode diagnostic for a resolved logarithmic transition with width $\Delta N=0.2$. The color scale denotes the exact numerical mode power divided by the smooth impulse approximation. The labeled contours indicate factors of two and separate the controlled impulse regime from the nonlinear resolved mass-ramp region where the $q_N^2$ contribution becomes dynamically significant.}
\label{fig:largePiCheck}
\end{figure}

The small-impulse analysis tests the profile-transform approximation itself. To probe the regime beyond linear response, we additionally integrate the full mode equation while retaining the nonlinear contribution proportional to $q_N^2$. Figure~\ref{fig:largePiCheck} shows the ratio between this exact finite-duration mode evolution and the smooth impulse approximation for the representative width $\Delta N=0.2$.

The two descriptions remain consistent throughout the controlled impulse regime relevant for the modes dominating the relic-abundance integral. However, once the resolved transition enters the regime $\Pi\gtrsim1$, the nonlinear contribution proportional to $q_N^2$ becomes dynamically important and produces a genuinely nonlinear mass-ramp effect.

For this reason, the relic-density-saturating calculations presented in the $r_A=1$ parameter space use $\Pi_{\rm DM}$ as the effective matching impulse inherited from the sharp non-slow-roll construction, while the resolved-profile analysis identifies the microphysical conditions required for a fully finite-duration completion of the transition.

The gravitational-wave calculation provides an additional consistency test of the source-level approximation. The fitted tensor spectrum of Eq.~\eqref{eq:GW_fit_sharp_main} reaches its maximum near
$x_*=f/f_*\simeq7.14$.
The finite-width source integration shown previously in Fig.~\ref{fig:gwshape} preserves the same peak normalization while introducing the finite-width filter on both vector spectra prior to the transverse-traceless convolution.

To test whether this source-level treatment accurately captures the full tensor evolution, we further insert the filtered spectra directly into an explicit radiation-era retarded Green-function integration while preserving the same tensor geometry.

\begin{figure}[htbp]
\centering
\includegraphics[width=\linewidth]{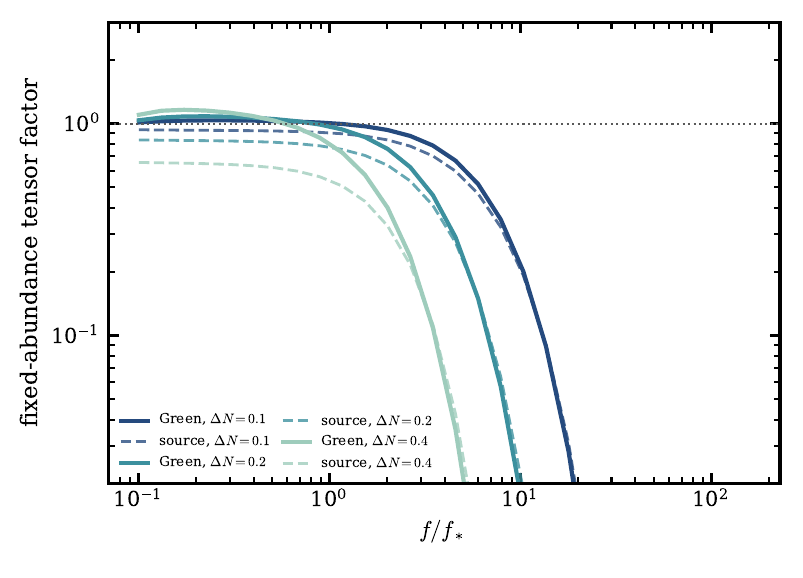}
\caption{Finite-width tensor calculation including an explicit radiation-era retarded Green-function integration. Solid curves show the fully time-integrated tensor calculation including the Green kernel and radiation-era transfer functions, while dashed curves retain only the source convolution. Both calculations incorporate the same finite-width filtering on the two vector spectra after imposing the relic abundance constraint.}
\label{fig:greenTensorCheck}
\end{figure}

Figure~\ref{fig:greenTensorCheck} compares the fully time-integrated Green-function calculation with the simpler source-only transverse-traceless approximation. The Green-function kernel leaves the near-peak normalization largely unchanged, confirming the validity of the source-level treatment near the dominant frequency scale. At higher frequencies, however, the full time integration produces a sharper ultraviolet cutoff once both vector legs become sensitive to the internal transition structure.

This final consistency check shows that the finite-width corrections remain stable after incorporating the complete radiation-era tensor evolution. The combined numerical calculations identify source-level finite-time effects in both the relic abundance and the induced gravitational-wave signal and support the analytic framework across the parameter ranges studied here.

\section{Physical interpretation and comparison with existing literature}\label{sec:interpretation}

\subsection{Connection with the original vector dark matter production mechanism}

In the original inflationary vector dark matter scenario, the final relic abundance is determined primarily by the vector mass $m$ and the inflationary Hubble scale $H_I$, once the longitudinal spectrum generated during inflation is evolved through the radiation dominated epoch until the vector becomes nonrelativistic \cite{GrahamMardonRajendran2016}. Within this framework, the mass scale required to reproduce the observed dark matter abundance remains strongly correlated with the inflationary scale, such that high scale inflation typically favors masses relevant for hidden photon searches outside the ultralight or fuzzy dark matter regime.

The subsequent non-slow-roll extension modifies this connection by allowing the effective vector mass during inflation to evolve independently from its late-time physical mass. As a consequence, ultralight vector masses can coexist with large inflationary fluctuations while maintaining the observed relic abundance \cite{LaRosaTasinato2025}.

The sharp mass excursion establishes the enhanced longitudinal production through the integrated impulse \cite{LaRosaTasinato2025}. That quantity fixes the zero frequency content of the event. The finite duration calculation determines the remaining finite frequency response. At fixed $\Pi$, profiles with different $\Delta N$ have the same asymptotic masses and the same coherent infrared response. Their response functions separate once $x\Delta N$ approaches unity because the vector mode accumulates phase while the mass is evolving. The profile transform $\FD$ then becomes an independent dynamical object. It fixes the resolved ultraviolet envelope while preserving the blue infrared branch. This gives a direct literature contrast. The sharp calculation determines the integrated impulse. The present calculation determines how a physical transition with that impulse is resolved by the momentum modes.

The distinction propagates differently into the two late observables. The relic abundance depends on the weighted one spectrum moment $\AD$, while the tensor source depends on the two spectrum convolution $\CD$. Enforcing the observed relic abundance changes the required impulse through $\AD$. The tensor prediction retains the ratio $\RD=\CD/\AD^2$. Two resolved transitions with the same relic abundance can therefore give different tensor responses when their finite frequency filters differ. The duration and differentiability of the transition carry information that survives abundance normalization. This is why the resolved profile is required in addition to the integrated impulse. The connection parallels scalar inflationary feature calculations in which modes that cross during an active background transition retain information about its time dependence \cite{Starobinsky1992,JacksonAssadullahiGowKoyamaVenninWands2024}.

\subsection{Connection with ultralight scalar and vector dark matter phenomenology}

The mass interval

\begin{equation}
10^{-24}{\rm eV}\lesssim m\lesssim10^{-18}{\rm eV}
\end{equation}

considered here overlaps directly with the parameter region commonly associated with ultralight scalar dark matter and fuzzy dark matter scenarios \cite{Hui2017,Hui2021,Eberhardt2025,Kobayashi2017}. Despite this overlap in mass scale, the underlying production mechanism differs fundamentally.

In conventional fuzzy dark matter models, the dark matter component is typically described as a coherently oscillating scalar condensate whose wave pressure modifies structure formation below a characteristic Jeans scale. By contrast, the inflationary vector mechanism considered here originates from quantum production of longitudinal vector modes during inflation. The resulting dark matter population carries intrinsic spin degrees of freedom, nontrivial polarization structure, and an initial spectrum qualitatively different from the scalar case.

Recent studies of ultralight vector dark matter have shown that anisotropic stress, polarization effects, and coherent metric oscillations can generate observational signatures distinct from scalar fuzzy dark matter, particularly in structure formation and pulsar timing measurements \cite{ChaseLeizerovichNacirLandau2024,ChaseCLASS2025,DrorWei2025}. Other coherent-vector constructions emphasize signatures associated with statistical anisotropy and model-dependent isocurvature transfer \cite{Nakayama2019,Nakayama2020Constraint,KitajimaNakayama2023}.

The framework developed here provides a concrete inflationary origin for such ultralight vector populations while simultaneously establishing a finite-duration consistency relation connecting relic abundance and induced gravitational-wave production.

\subsection{Relation to primordial black hole formation and non-attractor inflation}

Temporary departures from slow-roll inflation are widely employed to amplify small-scale curvature perturbations, thereby generating conditions favorable for primordial black hole formation as well as scalar-induced gravitational waves \cite{Sasaki2018,CarrKuhnel2020,ByrnesColePatil2019,Ragavendra2021,Domenech2021}.

The vector production mechanism studied here differs conceptually since it does not require direct enhancement of the adiabatic curvature perturbation itself. However, if the effective vector mass depends on the same inflationary background responsible for the departure from slow roll, then both scalar and vector sectors may be influenced by the same dynamical event.

This common origin makes finite transition duration physically relevant. A finite nonattractor phase defines the momentum range for efficient vector production and therefore determines the characteristic frequency window of the associated induced gravitational wave signal. The transition profile therefore enters the predicted spectrum through the resolved mode response and carries information in addition to the integrated impulse.

The profile-transform formalism developed in this work makes this connection particularly transparent. The scalar sector, vector sector, and induced tensor sector can all be written schematically as transfer kernels convolved with a finite-duration source. What differs between the sectors is the transfer kernel itself, whereas the differentiability properties of the source remain common.

A fully consistent microscopic model simultaneously solving the inflaton dynamics, vector mass evolution, curvature perturbations, vector abundance, and induced gravitational-wave production would replace the analytic profile function $S_\Delta$ by the exact numerically determined inflationary background. The formalism presented here then provides a direct diagnostic tool, since the effective finite-width envelope $\FD$ is simply the Fourier transform of the measured background transition profile.

\subsection{Possible microscopic origins of finite-duration transitions}

The transition profile $S_\Delta$ describes a localized dynamical event during inflation.

Several inflationary mechanisms naturally generate such finite-duration transitions. Inflection-point potentials and punctuated inflation scenarios can induce transient departures from slow roll within single-field inflation. In multifield models, rapid turns in field space can produce localized modifications of the effective adiabatic theory and significantly amplify fluctuations without requiring discontinuous background evolution \cite{Achucarro2012,PalmaSypsas2020}.

String-motivated inflationary models containing modulated potentials provide another natural realization, where oscillatory or localized features emerge in effective mass parameters, with the detailed envelope determined by heavy-field dynamics and microscopic modulations \cite{FlaugerMcAllisterPajerWestphalXu2009}.

Related examples arise in nonminimal dilaton inflation, where anomaly driven heavy sector effects generate calculable deformations of the inflationary background \cite{Pirzada:2026uak}. Resonant axion production mechanisms provide a complementary realization in which finite mass modulation directly controls the particle production process itself and directly changes the production history \cite{Pirzada:2026npl}.

The same physical logic appears in charged inflaton and symmetry breaking realizations of inflationary vector production, where the vector mass history follows a finite background excursion in place of an idealized discontinuous matching surface \cite{FirouzjahiGorjiMukohyamaSalehian2021,SalehianGorjiFirouzjahiMukohyama2021}. In Higgs realizations with a dynamical radial mode, the additional scalar evolution can modify the vector spectrum and abundance \cite{SatoTakahashiYamada2022,RediTesi2022}. The Stueckelberg realization specified in Sec.~\ref{sec:setup} describes the three polarization regime in which the background dependent Stueckelberg coefficient generates the mass history.

The hyperbolic tangent profile adopted in this work therefore captures the minimal physical information required by the problem: the integrated impulse strength, the transition duration measured in e-folds, and the differentiability properties of the underlying background evolution.

\subsection{Regime of validity and theoretical consistency}

The theoretical validity of the present framework is controlled by the mass generation assumptions stated in Sec.~\ref{sec:setup} together with five conditions governing vector production and late time transfer.

The calculation applies to a spectator vector throughout inflation, to the light-vector hierarchy in Eq.~\eqref{eq:light_hierarchy} for the modes dominating the abundance integral, and to sufficiently rapid reheating for the standard radiation-era transfer functions. The compact relic-density-saturating expressions describe the impulse-dominated branch $r_A=1$, while the gravitational-wave amplitude uses finite-width tensor kernels matched consistently to the published radiation-era vector spectrum.

The first three conditions coincide with those assumed in the original sharp-transition treatment. The fourth condition is enforced by Eq.~\eqref{eq:PiDM}, while the fifth is illustrated explicitly in Figs.~\ref{fig:tensorcheck} and \ref{fig:freqamp}.

The spectator constraint itself has two distinct components.

The first concerns the total energy density stored in the produced vector population,

\begin{equation}
 \rho_A^{\rm tr}\ll3\Mpl^2H_I^2,
\label{eq:backreaction_condition}
\end{equation}

which defines the dimensionless population parameter

\begin{equation}
 \epsilon_A^{\rm pop}\equiv\frac{\rho_A^{\rm tr}}{3\Mpl^2H_I^2}
 \simeq \frac{H_I^2}{12\pi^2\Mpl^2}\left(\frac12+\Pi+\Pi^2\AD\right).
\label{eq:spectator_parameter}
\end{equation}

\begin{figure}[htbp]
\centering
\includegraphics[width=9cm, height=8cm]{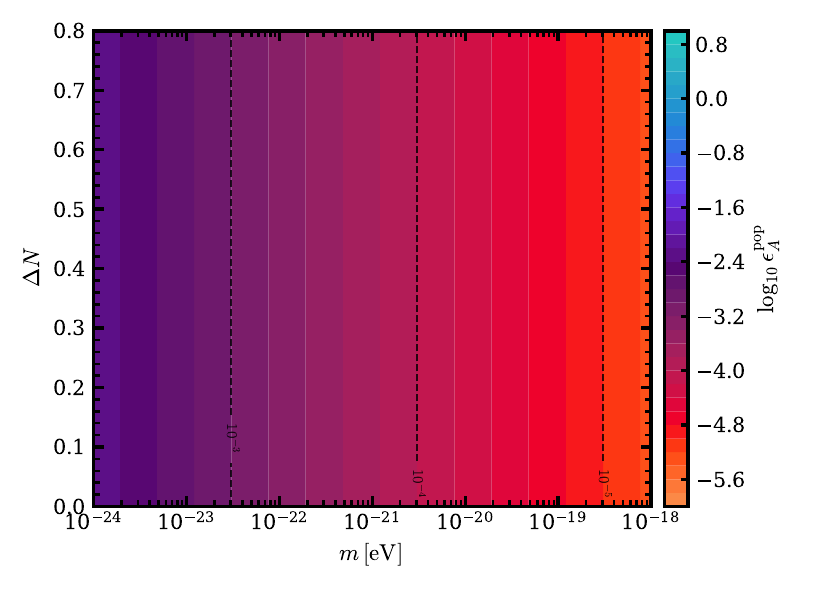}
\caption{Spectator-sector population constraint for the relic-density-saturating case $r_A=1$ at $H_I=10^{14}{\rm GeV}$. Contours indicate the parameter $\epsilon_A^{\rm pop}$ after imposing the abundance constraint. The physically allowed spectator regime corresponds to $\epsilon_A^{\rm pop}\ll1$.}
\label{fig:spectatorparameter}
\end{figure}

Figure~\ref{fig:spectatorparameter} shows the resulting constraint for the relic-density-saturating case $r_A=1$ assuming $H_I=10^{14}{\rm GeV}$.

A second and potentially stronger constraint arises from derivative-sensitive stress generated during the transition itself. Since the finite transition satisfies

\begin{equation}
 q_N=1-\Pi S_\Delta'(N),\qquad q_{NN}=-\Pi S_\Delta''(N),
\end{equation}

the instantaneous stress associated with the sector generating $J(N)$ can scale parametrically as $\Pi/\Delta N$ and $\Pi/\Delta N^2$.

A conservative dimensionless estimate of this transition-induced stress is

\begin{equation}
 \epsilon_A^{J}\equiv
 \frac{H_I^2}{12\pi^2\Mpl^2}\left(\frac{\Pi_{\rm DM}}{\Delta N}\right)^2,
\label{eq:derivative_parameter}
\end{equation}

which quantifies the parameter region where the transition-induced stress approaches the background inflationary energy density.

\begin{figure}[htbp]
\centering
\includegraphics[width=10cm, height=8cm]{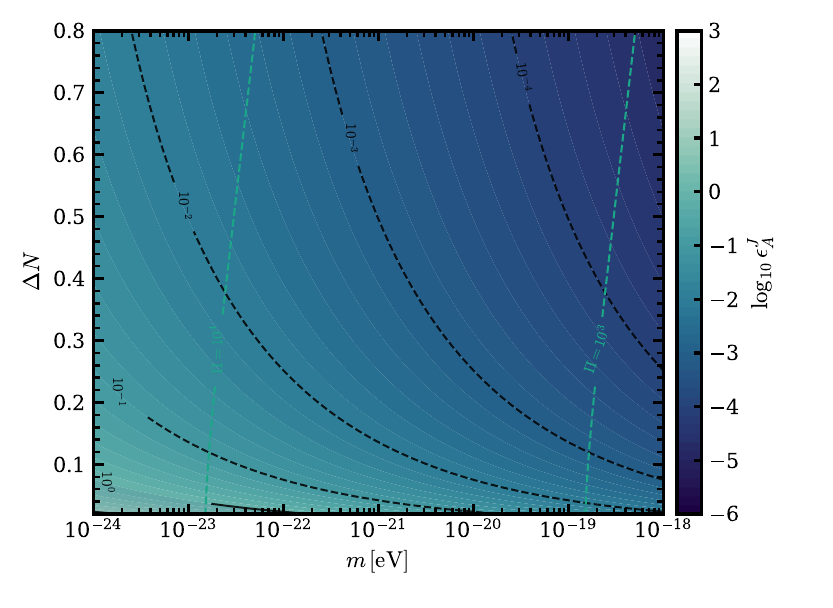}
\caption{Derivative-sensitive transition-stress constraint evaluated for $H_I=10^{14}{\rm GeV}$. The quantity $\epsilon_A^J$ estimates the stress scale associated with varying $J$ over a finite interval $\Delta N$. The figure identifies the region where the sector generating the time-dependent vector mass contributes non-negligibly to the inflationary energy budget.}
\label{fig:derivativeparameter}
\end{figure}

Figure~\ref{fig:derivativeparameter} demonstrates that this derivative-sensitive constraint becomes more restrictive than the population constraint in the regime of small transition width and low vector mass.

The coefficient appearing in Eq.~\eqref{eq:derivative_parameter} should be interpreted as a conservative normalization estimate. In the Stueckelberg realization of Sec.~\ref{sec:setup}, a fully specified trigger potential would replace this estimate with the exact stress energy contribution of $\chi$ during the motion of its inflation dependent minimum.

Finally, large-scale isocurvature constraints remain naturally suppressed since finite-duration effects do not modify the characteristic blue infrared spectrum. Measurements of primordial non-adiabatic perturbations by the \cite{PlanckInflation2018,PlanckParameters2018} collaboration strongly constrain isocurvature contributions on CMB scales.

Throughout this analysis, the transition scale $k_1$ is assumed to remain well separated from observable CMB scales, analogous to the scale separation commonly employed in primordial black hole motivated non-slow-roll inflationary scenarios.

Smooth filtering of early-time inflationary memory appears in related studies of inflationary initial states, where prolonged expansion confines residual information to a finite comoving region \cite{Khan:2026doo}. If the transition were shifted toward observable CMB scales, a complete treatment would necessarily include metric perturbations together with explicit cross-correlations involving the curvature sector. The central conceptual result of this work is that finite-duration inflationary transitions introduce an independent physical timescale that survives both relic-abundance fixing and tensor-source evolution, thereby promoting the internal structure of inflationary background transitions into a directly observable component of ultralight vector dark matter phenomenology.

\section{Conclusions}

A finite-duration non-slow-roll phase replaces the instantaneous mass transition by a resolved momentum filter for longitudinal vector production. This construction preserves the enhanced ultralight-vector branch while determining how transition smoothing enters the relic abundance and the induced gravitational-wave signal.

The central result of our analysis is that the effects of a finite-duration transition are governed by the Fourier transform of the transition profile. For a smooth hyperbolic tangent transition with duration $\Delta N$ e-folds, the singular sharp-transition impulse is replaced by the finite-width transfer function
$\FD(x)={\pi \Delta N x/2}/{\sinh(\pi \Delta N x/2)}$.
This profile-dependent kernel provides a direct physical measure of how the finite temporal structure of the background controls the excitation efficiency of individual momentum modes. In particular, modes with wavelengths larger than the transition scale remain unaffected, whereas shorter-wavelength modes capable of resolving the transition are progressively suppressed.

We showed that this finite-time effect preserves the infrared structure responsible for efficient vector production while eliminating the ultraviolet oscillatory behavior characteristic of the idealized sudden-transition approximation. The resulting relic abundance is determined by the weighted spectral moment $\AD=\langle\FD^2\rangle_W$, which quantifies the suppression of the produced vector population once the finite transition profile is taken into account. This establishes that the enhanced abundance branch identified in the sharp-transition approximation remains physically stable under realistic smoothing, although the required transition impulse is modified quantitatively.

The induced tensor signal exhibits a more intricate dependence. Unlike the relic abundance, which depends on a single spectral moment, the gravitational-wave source is generated through a convolution of two filtered vector spectra, introducing an independent tensor-source correction factor. After imposing a fixed relic abundance, the peak gravitational-wave amplitude satisfies the relation
$\Omega_{\rm GW}^{\rm max}\simeq1.2\times10^{-10}\,\rDM^2
\left(\frac{1.5\times10^{-21}{\rm eV}}{m}\right)\RD$,
where $\RD=\CD/\AD^2$ encodes the residual source-level dependence arising from finite-width tensor kernels. This demonstrates that abundance suppression alone does not fully determine the gravitational-wave prediction, since finite transition duration simultaneously reshapes the tensor production kernel itself.

Our numerical analysis confirms the validity of the analytic finite-width formalism. Direct integration of the longitudinal mode equation reproduces the predicted transfer envelope with percent-level agreement over the momentum region dominating relic production, while explicit radiation-era tensor calculations verify the source-kernel dependence entering the induced gravitational-wave amplitude. These consistency checks establish that the profile-transform formalism provides an accurate description of the finite-duration non-slow-roll dynamics over the physically relevant parameter space.

The duration and differentiability of the departure from slow roll provide dynamical information in addition to the integrated mass impulse. A gauge invariant Stueckelberg coefficient controlled by a transient heavy field background generates the mass history used in the calculation with three massive vector polarizations. A Higgs realization with an active radial mode contains an additional cosmological degree of freedom whose evolution enters the vector production problem. A microscopic inflationary realization fixes both the integrated excursion and its temporal structure, and these two pieces enter the late time abundance and induced tensor background through different profile functionals.

For the ultralight vector mechanism considered here, sub-e-fold transitions generate order-one changes in the abundance and gravitational-wave predictions while smoothing the ultraviolet ringing of singular matching. The finite-duration framework thereby connects microscopic non-slow-roll dynamics with ultralight dark matter abundance and low-frequency primordial gravitational-wave observables.

\appendix

\section{Derivation of the longitudinal Proca action}\label{app:proca}

In conformal coordinates, the action given in Eq.~\eqref{eq:proca_action} takes the form  
\begin{equation}
 S_A=\int\dd\tau\dd^3x\left[-\frac14F_{\mu\nu}F_\eta^{\mu\nu}-\frac12m^2J^2\eta^{\mu\nu}A_\mu A_\nu\right],
\end{equation}
where the tensor $F_\eta^{\mu\nu}$ is defined with indices raised using the Minkowski metric. Decomposing the spatial vector field into transverse and longitudinal components according to $A_i=A_i^T+\partial_i\varphi$, the longitudinal contribution in Fourier space becomes  
\begin{equation}
\begin{split}
S_L=\frac12\int\dd\tau\dd^3k\Big[&k^2(A_{0,k}-\varphi_k')(A_{0,-k}-\varphi_{-k}') \\
&+m^2J^2A_{0,k}A_{0,-k}-m^2J^2k^2\varphi_k\varphi_{-k}\Big].
\end{split}
\label{eq:app_action_A0}
\end{equation}

The nondynamical field $A_{0,k}$ is determined through its constraint equation. Varying the action with respect to $A_{0,-k}$ yields  
\begin{equation}
 k^2(A_{0,k}-\varphi_k')+m^2J^2A_{0,k}=0,
\end{equation}
which directly reproduces Eq.~\eqref{eq:A0_constraint}. Substituting this expression back into the action leads to Eq.~\eqref{eq:long_action_phi}. Introducing the canonical field redefinition $\pi_k=Z_k\varphi_k$, where $Z_k$ is defined in Eq.~\eqref{eq:Zdef}, the action can be recast as  
\begin{equation}
 S_L=\frac12\int\dd\tau\dd^3k\left[\pi_k'\pi_{-k}'-\left(k^2+m^2J^2-\frac{Z_k''}{Z_k}\right)\pi_k\pi_{-k}\right]
\end{equation}
after performing an integration by parts.

The time-dependent normalization factor satisfies  
\begin{equation}
 \frac{Z_k''}{Z_k}=\frac{k^2}{k^2+m^2J^2}\frac{J''}{J}-\frac{3k^2m^2J'^2}{(k^2+m^2J^2)^2},
\end{equation}
from which the effective frequency expression in Eq.~\eqref{eq:omega_general} follows immediately.

\section{Conformal-time identities}\label{app:time}

For an exact de Sitter background, the scale factor evolves as  
\begin{equation}
 a=-\frac{1}{H_I\tau}.
\end{equation}
Defining the e-fold variable $N=\ln(a/a_1)$, differentiation gives  
\begin{equation}
 \frac{\dd N}{\dd\tau}=-\frac{1}{\tau}.
\end{equation}
The corresponding conformal-time derivatives can therefore be expressed as  
\begin{equation}
 \partial_\tau=-\tau^{-1}\partial_N,
\qquad
 \partial_\tau^2=\tau^{-2}(\partial_N^2+\partial_N).
\end{equation}

For a time-dependent function written as $J=\exp q(N)$, derivatives with respect to $N$ satisfy  
\begin{equation}
 J_N=q_NJ,
\qquad J_{NN}=(q_{NN}+q_N^2)J.
\end{equation}
Combining these relations yields Eq.~\eqref{eq:Jidentity}. For the specific transition profile defined by $q=N-\Pi S_\Delta(N)$, one obtains  
\begin{equation}
 q_N=1-\frac{\Pi}{2\Delta N}\mathrm{sech}^2\left(\frac{N}{\Delta N}\right),
\end{equation}
and  
\begin{equation}
 q_{NN}=\frac{\Pi}{\Delta N^2}\mathrm{sech}^2\left(\frac{N}{\Delta N}\right)\tanh\left(\frac{N}{\Delta N}\right).
\end{equation}
These expressions are used directly in the numerical integration of the longitudinal mode equation.

\section{Fourier transform of the tanh step}\label{app:fourier}

The normalized derivative associated with the smooth hyperbolic tangent transition is given in Eq.~\eqref{eq:Sprime}. Introducing the rescaled variable $y=N/\Delta N$, the profile transform can be written as  
\begin{equation}
 \FD(x)=\frac12\int_{-\infty}^{\infty}\dd y\,\mathrm{sech}^2y\,\e^{\ii x\Delta N y}.
\end{equation}

This integral is evaluated using the standard contour identity  
\begin{equation}
 \int_{-\infty}^{\infty}\dd y\,\mathrm{sech}^2y\,\e^{\ii qy}=\frac{\pi q}{\sinh(\pi q/2)}.
\end{equation}
Substituting this result immediately gives Eq.~\eqref{eq:formfactor_main}. The small-argument expansion follows from expanding the ratio $z/\sinh z$, where $z=\pi\Delta N x/2$, yielding  
\begin{equation}
 \frac{z}{\sinh z}=1-\frac{z^2}{6}+\frac{7z^4}{360}-\frac{31z^6}{15120}+O(z^8).
\end{equation}

\section{Abundance moment and analytic checks}\label{app:moment}

The normalization factor appearing in Eq.~\eqref{eq:AD_main} remains finite and admits a simple analytic evaluation,  
\begin{equation}
 \int_0^\infty\dd\ln x\,\frac{x^2}{(1+x^2)^2}
 =\int_0^\infty\frac{x\,\dd x}{(1+x^2)^2}=\frac12.
\end{equation}
The abundance suppression factor therefore reduces to  
\begin{equation}
 \AD=2\int_0^\infty\frac{x\,\dd x}{(1+x^2)^2}\FD^2(x).
\end{equation}

To examine the small-width limit, one may insert the perturbative expansion given in Eq.~\eqref{eq:Fsmall}, leading formally to  
\begin{equation}
 \AD=1-\frac{\pi^2\Delta N^2}{12}\frac{\int x^3(1+x^2)^{-2}\dd x}{\int x(1+x^2)^{-2}\dd x}+\cdots.
\end{equation}

However, this expansion is not uniformly convergent over the full momentum range, since the integral involving $x^3/(1+x^2)^2$ develops logarithmic sensitivity to the ultraviolet region. Consequently, although $\AD(\Delta N)$ varies smoothly in the numerical calculation, a low-order Taylor approximation does not accurately reproduce its global behavior. The abundance integral therefore retains the complete transform.

\section{Fixed-abundance tensor relation}\label{app:relation}

Starting from Eq.~\eqref{eq:smooth_abundance_main}, we define the shorthand quantity  
\begin{align}
 B&=10^{-7}\left(\frac{m}{m_0}\right)^{1/2}
 \left(\frac{H_I}{H_0}\right)^2,\\
 m_0&=1.5\times10^{-21}{\rm eV},
 \qquad H_0=10^{14}{\rm GeV}.
\end{align}

The relic abundance can then be expressed as  
\begin{equation}
 \rDM=B\left(\frac12+\Pi+\Pi^2\AD\right).
\end{equation}

Assuming the gravitational-wave source at fixed impulse takes the form  
\begin{align}
 \Omega_{\rm GW}^{\rm max}(\Pi,\Delta N)
 &=1.2\times10^{-24}
 \left(\frac12+\Pi+\Pi^2\AD\right)^2 \nonumber\\
 &\quad\times
 \left(\frac{H_I}{H_0}\right)^4\RD ,
\end{align}
the abundance factor can be eliminated algebraically, leading directly to Eq.~\eqref{eq:master_relation_main},  
\begin{equation}
 \Omega_{\rm GW}^{\rm max}=1.2\times10^{-10}\,\rDM^2
 \left(\frac{m_0}{m}\right)\RD.
\end{equation}

The residual factor $\RD$ becomes unity only when the finite-width modification of the tensor source scales identically to the square of the abundance moment. In contrast, a fixed-impulse comparison preserves $\Pi$ and isolates the direct suppression associated with the tensor convolution factor $\CD$.

\section{Numerical mode-envelope check}\label{app:modecheck}

In the light-vector limit, the canonical longitudinal mode satisfies the evolution equation given in Eq.~\eqref{eq:mode_N_numeric}. Initial conditions are imposed at an early time $N_i$ using the positive-frequency de Sitter solution, where $y=k|\tau|=xe^{-N}$. The mode and its derivative are therefore initialized as  
\begin{align}
 \pi_k(N_i)&=\left(1+\frac{i}{y_i}\right)e^{iy_i},\\
 \pi_{k,N}(N_i)&=-y_i\frac{\partial}{\partial y_i}
 \left[\left(1+\frac{i}{y_i}\right)e^{iy_i}\right].
\end{align}

Since only the relative response is relevant, the overall normalization cancels in the final ratio. The transition profile enters through the background derivatives  
\begin{equation}
 q_N=1-\frac{\Pi}{2\Delta N}{\rm sech}^2\left(\frac{N}{\Delta N}\right),
\end{equation}
and  
\begin{equation}
 q_{NN}=\frac{\Pi}{\Delta N^2}{\rm sech}^2\left(\frac{N}{\Delta N}\right)
 \tanh\left(\frac{N}{\Delta N}\right).
\end{equation}

Within the linear-response regime, the late-time ratio defined in Eq.~\eqref{eq:mode_envelope_ratio} removes the universal slow-roll contribution and isolates the finite-width correction associated with the transition profile. The numerical solutions shown in Fig.~\ref{fig:modecheck} confirm that the analytic transform reproduces the expected suppression envelope across the momentum interval dominating the relic abundance integral. The remaining discrepancies are largest in the high-momentum regime where both analytic and numerical mode amplitudes are already exponentially suppressed.

\bibliographystyle{apsrev4-2}
\bibliography{references}
\end{document}